\documentclass[keywords,pdf]{iucr}              
\usepackage{amsmath}
\usepackage{graphicx}
\usepackage{amssymb}
\usepackage{layouts}

\begin{document}  

\title{Ptychographic X-ray Speckle Tracking}
\shorttitle{Speckle tracking}

\cauthor[a,b]{Andrew J.}{Morgan}{morganaj@unimelb.edu.au}{}
\author[a]{Harry M.}{Quiney}
\author[c]{Sa\v{s}a}{Bajt}
\author[b,d,e]{Henry N.}{Chapman}

\aff[a]{ARC Centre of Excellence in Advanced Molecular Imaging, School of Physics, University of Melbourne, Parkville, Victoria 3010, Australia}
\aff[b]{CFEL, Deutsches Elektronen-Synchrotron DESY, 
Notkestra{\ss}e 85, 22607 Hamburg, Germany}
\aff[c]{DESY, Notkestra{\ss}e 85, 22607 Hamburg, Germany}
\aff[d]{Department of Physics, University of Hamburg, 
Luruper Chaussee 149, 22761 Hamburg, Germany}
\aff[e]{Centre for Ultrafast Imaging, Luruper Chaussee 149, 22761 Hamburg, Germany}

\shortauthor{Andrew J. Morgan \textit{et al.}}


\maketitle 

\begin{synopsis}
Here we present method for the simultaneous measurement of a wavefront's phase and projection hologram of an unknown sample. This method relies on an updated form of the speckle tracking approximation, which is based on a second order expansion of the Fresnsel integral.
\end{synopsis}

\begin{abstract}
We present a method for the measurement of the phase gradient of a wavefront by
tracking the relative motion of speckles in projection holograms as a sample
is scanned across the wavefront. By removing the need to obtain an un-distorted
reference image of the sample, this method is suitable for the metrology of
highly divergent wavefields. Such wavefields allow for large magnification 
factors, that, according to current imaging capabilities, will allow for 
nano-radian angular sensitivity and nano-scale sample projection imaging. 
Both the reconstruction algorithm and the imaging geometry are nearly 
identical to that of ptychography, except that the sample is placed downstream
of the beam focus and that no coherent propagation is explicitly accounted
for. Like other x-ray speckle tracking methods, it is robust to low-coherence
x-ray sources making is suitable for lab based x-ray sources. Likewise it is
robust to errors in the registered sample positions making it suitable for 
x-ray free-electron laser facilities, where beam pointing fluctuations can be
problematic for wavefront metrology. We also present a modified form of the 
speckle tracking approximation, based on a second-order local expansion of the 
Fresnel integral. This result extends the validity of the speckle tracking 
approximation and may be useful for similar approaches in the field.
\end{abstract}

\section{Introduction}



New facilities are providing ever more brilliant x-ray sources. To access the 
full potential of these sources we need x-ray optics that are capable of 
focusing light to meet the requirements of various imaging modalities.  Thus there 
is an increasing need for at-wavelength and \textit{in-situ} wavefront 
metrology techniques that are capable of measuring the performance of these 
optics to the level of their desired performance. This is a challenging task, 
as current x-ray optics technologies are attaining focal spot sizes below 
$10\;$nm \cite{Huang2013, Mimura2010a, Morgan2015, Bajt2018, 
Murray2019a}. 
Furthermore, adaptive optics are being employed to correct 
for wavefront aberrations by altering the physical state 
of a lens system in response to real-time measurements of wavefront errors 
\cite{Mercere2006, Zhou2019}. Such systems therefore benefit from fast and 
accurate wavefront metrology for rapid feedback. 

Wavefront metrology techniques generally fall into one of three catagories 
\cite{Wilkins2014, Wang2015}: (i) direct phase measurements, such as interferometry 
using single crystals \cite{Bonse1965}; (ii) phase gradient measurements, such as 
Hartmann sensors \cite{Lane:92}, coded aperture methods \cite{Olivo2007}
and grating-based interferometry \cite{David2002}; and (iii) propagation-based 
methods sensitive to the secondary derivative of the wavefront's phase 
\cite{Wilkins1996, Wang2015, Berujon2014}. 

One such method, falling into the second category above, was introduced 
by \citeasnoun{Berujon2012}\footnote{A similar method, based on the same principles, 
was later developed by \citeasnoun{Morgan2012a} (no relation to the current author)}. 
This method is a wavefront metrology technique
based on near-field speckle-based imaging, which they term the ``x-ray speckle
tracking'' (XST) technique. In XST, the 2D phase gradient of
a wavefield can be recovered by tracking the displacement of localised 
``speckles'' between an image and a reference image produced in the projection hologram of an object with a random phase/absorption profile. Additionally, 
XST can be employed to measure the phase profile of an object's transmission 
function. Thanks to the simple experimental set-up, high angular sensitivity 
and compatibility with low coherence sources this method has since been 
actively developed for use in synchrotron and laboratory light sources, see 
\cite{Zdora2018} for a recent review. 

In ptychography, a sample is scanned across the beam wavefront (typically at or 
near the focal plane of a lens) while diffraction data is collected in the 
far-field of the sample. An iterative algorithm is usually employed to update 
initial estimates for the complex wavefront of the illumination and the sample transmission 
functions. If illuminated regions of the sample overlap sufficiently, 
then it is possible for a unique solution for both of these functions to be 
obtained \cite{Hue2010}.
Thus, ptychography is an imaging modality that performs both 
aberration free sample imaging and wavefront metrology simultaneously. This is 
in contrast to XST where these two imaging modalities correspond to separate 
imaging geometries. 

\onecolumn
\begin{table}
    \caption{The PXST method}
\scalebox{1.0}{\begin{tabular}{lcl}
\\ [-0.2cm] 
\textbf{Governing equation} &  
$I_n(\textbf{x}, z) \approx W(\textbf{x}) I_\text{ref}(\textbf{u}(\textbf{x})- \Delta \textbf{x}_n, \bar{z})$ 
& See section \ref{sec:method} and section \ref{sec:approx} of the appendix.\\ [0.2cm]
 & $ w(\textbf{x}) \left( \frac{\bar{z}}{z}\right)^2 I_\text{ref}(\textbf{x}, \bar{z}) \approx   I_n(\textbf{u}^{-1}(\textbf{x}+\Delta \textbf{x}_n), z)$ & Reciprocal form for the above equation. $\textbf{u}^{-1}$ is the inverse of $\textbf{u}$. \\ [0.2cm]
\textbf{Target function} & 
$\varepsilon = \sum_n \iint d\textbf{x}
    \frac{1}{\sigma^2_I(\textbf{x})} \left[I_n(\textbf{x}, z) -  W(\textbf{x}) 
    I_\text{ref}(u(\textbf{x})-\Delta \textbf{x}_n, \bar{z})
    \right]^2$  & Equation \ref{eq:calc_err} in section \ref{sec:algorithm}. To be minimised wrt $I_\text{ref}$, $\nabla \Phi$ and $\Delta \textbf{x}_n$.\\ [0.2cm] 
\textbf{Geometric mapping} & $\textbf{u}(\textbf{x}) = \textbf{x} - \frac{\lambda z}{2\pi} 
          \nabla \Phi(\textbf{x})$ & See Eq. \ref{eq:u} in appendix \ref{sec:approx}. \\[0.2cm]
 & $\textbf{u}^{-1}(\textbf{x}) = \textbf{x} + \frac{\lambda z}{2\pi} 
          \nabla \phi(\textbf{x})$ & Reciprocal form for the above equation. See Eq. \ref{eq:g} in appendix \ref{sec:approx}. \\[0.2cm]
\textbf{Imaging geometry} & see Fig. \ref{fig:overview} & Described in section \ref{sec:background}. \\ [0.2cm] 
\textbf{Iterative update algorithm} & see Fig. \ref{fig:alg} & Described in section \ref{sec:method}. \\ [0.2cm]
\textbf{Angular sensitivity} & $\Delta \Theta_\phi = \sigma_\text{eff} / z$ & In the plane of the sample, see Eq. \ref{eq:dtphi}. \\ [0.2cm]
                             & $\Delta \Theta_\Phi = \sigma_\text{eff} / (zM)$ & In the plane of the detector, see Eq. \ref{eq:dtPhi}. \\ [0.2cm]
\textbf{Phase sensitivity}   & $\Delta \phi = \Delta \Phi = \frac{2\pi}{\lambda} \frac{\sigma_\text{eff}^2}{zM} $ & Sample/Detector plane, see Eq. \ref{eq:dPhi}. \\ [0.2cm]
\end{tabular}}
\label{table:PXST}
\end{table}
\twocolumn

We propose a combined approach, which we term Ptychographic X-ray Speckle 
Tracking (PXST). In this approach, near-field inline holograms are recorded 
as an unknown sample is scanned across an unknown wavefield. Estimates for the 
un-distorted sample projection image and the wavefield are then updated based
on the observed speckle displacements. There is no reference image and no 
additional speckle-producing object is required. This imaging geometry allows 
for XST to be used for highly divergent x-ray beams, thus expanding the 
applicability of this simple and robust method to include next generation high 
numerical aperture x-ray lenses. 

\citeasnoun{Berujon2014} have proposed a similar method, also based on XST 
and compatible with highly divergent beams. 

In their approach, the second derivative of the wavefront phase is measured. 
Additionally, nano-radian angular sensitivity can be achieved with relatively small 
step sizes in the scan of the sample on a piezo-driven stage (discussed further in the next section). 
In contrast, PXST more closely aligns 
with current XST based techniques, such as the ``unified modulated pattern analysis'' method of \cite{Zdora, Zdora:18}, that do not rely on small sample translations.

In section \ref{sec:background}, we briefly review the XST method and its 
extension to PXST. In section \ref{sec:method} we present the governing 
equation, which is based on a second-order expansion of the Fresnel diffraction
integral (presented in section \ref{sec:approx}). The region of validity for 
the speckle tracking approximation determines the applicable imaging geometries, 
which are presented in section \ref{sec:limits}. 
We present the iterative reconstruction 
algorithm and the target function, which is to be minimised by the algorithm, in section \ref{sec:algorithm}. Conditions for the uniqueness of the
solution are discussed in section \ref{sec:uniqueness}. Finally, the theoretically achievable 
angular sensitivity of the wavefront reconstruction as well as the imaging 
resolution of the sample projection image are then presented in section 
\ref{sec:res}. For reference we define the mathematical symbols used throughout the paper in table \ref{table:symbols}.
In table \ref{table:PXST} we summarise the main results of this article and refer the reader to the relevant sections.

\begin{table}
    \caption{Symbols}
\begin{tabular}{cl}
$I_n(\textbf{x}, z)$ 
& n\textsuperscript{th} recorded image  \\ [0.1cm]
$I_\text{ref}(\textbf{x}, z)$ 
& reference projection image of the sample  \\[0.1cm]
$\Delta \textbf{x}_n$ 
& displacement of sample in transverse plane \\[0.1cm]
$T(\textbf{x})$ 
& transmission function of the quasi-2D sample \\[0.1cm]
$z_1$
&source-to-sample distance \\[0.1cm]
$z$ 
&sample-to-detector distance \\[0.1cm]
$\bar{z} = \frac{z z_1}{z_1 + z}$
&effective propagation distance \\[0.1cm]
$M = \frac{z_1+z}{z_1}$ 
& geometric magnification factor \\[0.1cm]
$\lambda$
& wavelength of radiation \\[0.1cm]
$\sigma_\text{eff}$
& smallest resolvable speckle displacement \\
& in the plane of the detector \\[0.1cm]
$i$ 
& imaginary number \\[0.1cm]
$\textbf{a} \cdot \textbf{b}$ 
& dot product between vectors $\textbf{a}$ and $\textbf{b}$ \\[0.1cm]
$p(\textbf{x}, 0) = \sqrt{w}(\textbf{x}) e^{i\phi(\textbf{x})}$
& illumination wavefront in the sample plane. \\ 
& $w$ and $\phi$ are the intensity and phase respectively. \\[0.1cm]
$p(\textbf{x}, z) = \sqrt{W}(\textbf{x}) e^{i\Phi(\textbf{x})}$
& illumination wavefront in the detector plane. \\
& $W$ and $\Phi$ are the intensity and phase respectively. \\[0.1cm]
$\textbf{x} \equiv (x, y)$
& transverse coordinate \\[0.1cm]
$\nabla \equiv (\frac{\partial }{\partial x}, \frac{\partial }{\partial y})$
& transverse gradient operator \\[0.1cm]
\end{tabular}
\label{table:symbols}
\end{table}

\section{Background}\label{sec:background}

The problem with wavefront metrology is that it is much more difficult to 
measure a wavefront's phase than its intensity; the intensity can be measured 
directly by placing a photon counting device at any point in the wavefront's 
path, whereas the phase information is indirectly encoded in the wavefront's 
intensity profile as it propagates through space. For plane wave illumination, 
no measurement of the wavefront's intensity alone will reveal its direction of 
propagation. One solution to this problem is to place an absorbing object at a 
known point in the path of the light from which the direction of propagation 
can then be inferred from the relative displacement between the centre of the 
object and the shadow cast on a screen some distance away, just as the angle of 
the sun can be estimated by following the line from a shadow to its object.

\begin{figure}
\includegraphics[width=8.88cm]{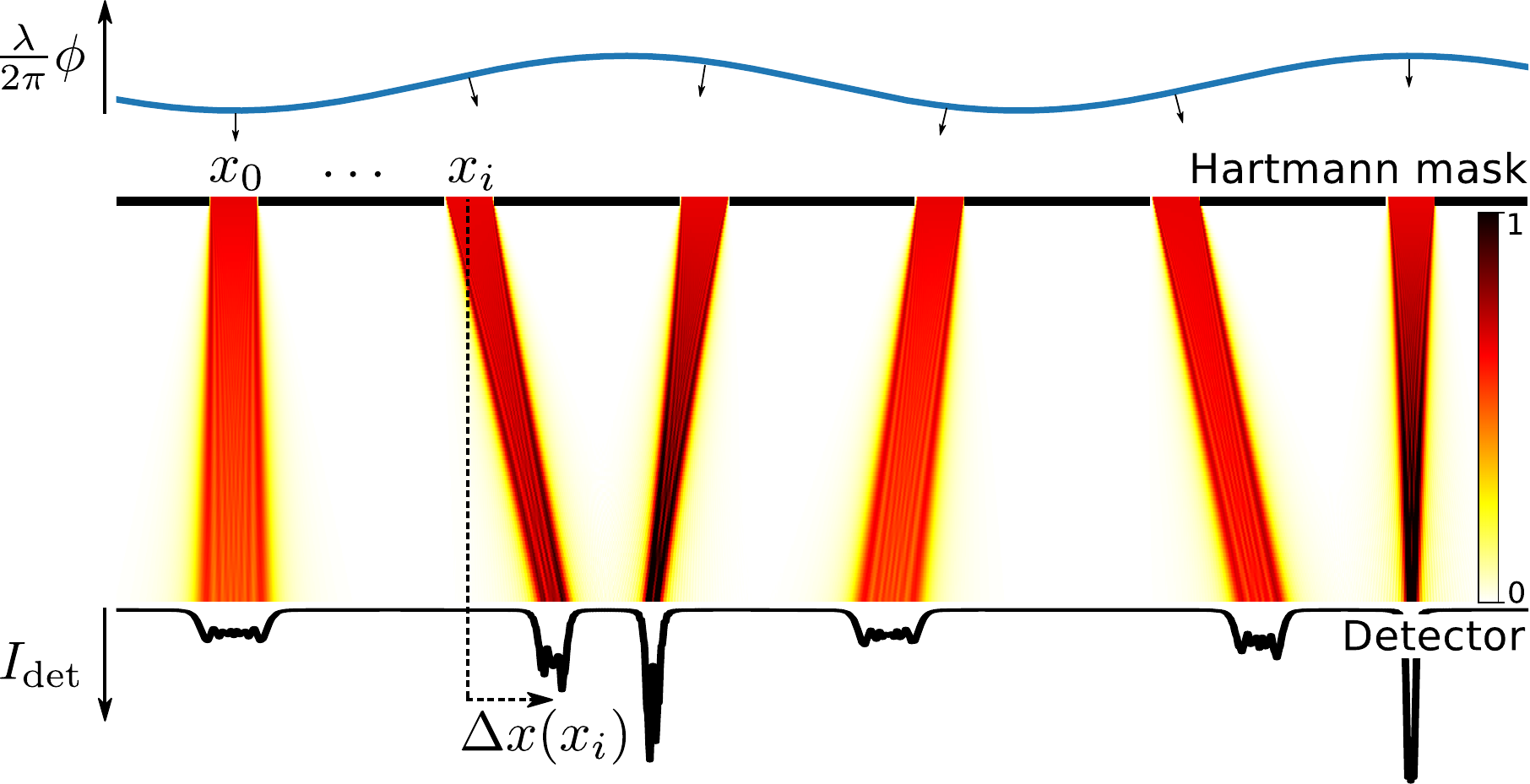}
\caption{Illustration of the Hartmann sensing principle. \textbf{Top:} 
    phase of the wavefront incident on the entrance surface of the mask. The 
    phase has been scaled by $\lambda / 2\pi$ so that the normal to the tangent 
    is parallel to the local direction of propagation of the wavefront. The 
    arrows indicate the direction of propagation at the centre of each mask 
    hole. \textbf{Middle:} intensity of the wavefront as it propagates from the 
    mask (top) to the detector (bottom). The colour scale is shown on the left. 
    \textbf{Bottom:} the one-dimensional intensity profile of the wavefront  
    as measured by the detector.}
\label{fig:Hartmann}
\end{figure}

This simple idea forms the basis of the Hartmann sensor \cite{Daniel1992}, shown 
schematically in Fig \ref{fig:Hartmann}. Originally designed to measure aberrations 
in telescopes and later for atmospheric distortions, the Hartmann sensor 
can be used as an x-ray wavefront metrology tool by cutting a regular grid of 
small holes, spaced at known intervals (say $x_i$ where $i$ is the hole index), 
in a mask and then recording the shadow image on a detector, which is placed 
a small distance downstream of the mask. 

Provided that each hole can be matched with each shadow image, the 
angle made between them, $\Theta(x_i) = \arctan{(\Delta x(x_i) / z)}$ (in 1D), is 
equal to the average direction of propagation of that part of the wavefront 
passing through each hole $\Theta(x_i) = \frac{\lambda}{2\pi d}  
\int_{x_i-d/2}^{x_i+d/2} \frac{\partial \phi(x)}{\partial x}dx $, where 
$\Delta x(x_i)$ is the observed displacement along $x$ of the $i$\textsuperscript{th} shadow, 
$z$ is the distance between the mask and the detector and $d$ is the hole width. 
With a suitable interpolation routine, $\Theta(x)$ 
can be estimated from the set of $\Theta(x_i)$ and the phase profile can be 
obtained up to a constant with
\begin{align} \label{eq:int_angle}
   \phi(x) &\approx \frac{2\pi}{\lambda} \int \Theta(x)\; dx.
\end{align}

One limitation of this technique is that the resolution obtained is limited by 
the spacing between each hole in the mask. For example,  \citeasnoun{Mercere2006} 
used a Hartmann sensor in an active optic system with a 
grid of $75\times 75$ square holes over a $10\;\text{mm} \times 10\;\text{mm}$ 
area, whereas the CCD detector had a $1024\times 1024$ grid of pixels over a 
$13\;\text{mm} \times 13\;\text{mm}$ area. Thus the 
Hartman sensor had a resolution 10.5 times worse than the CCD detector.

\begin{figure}
\includegraphics[width=8.88cm]{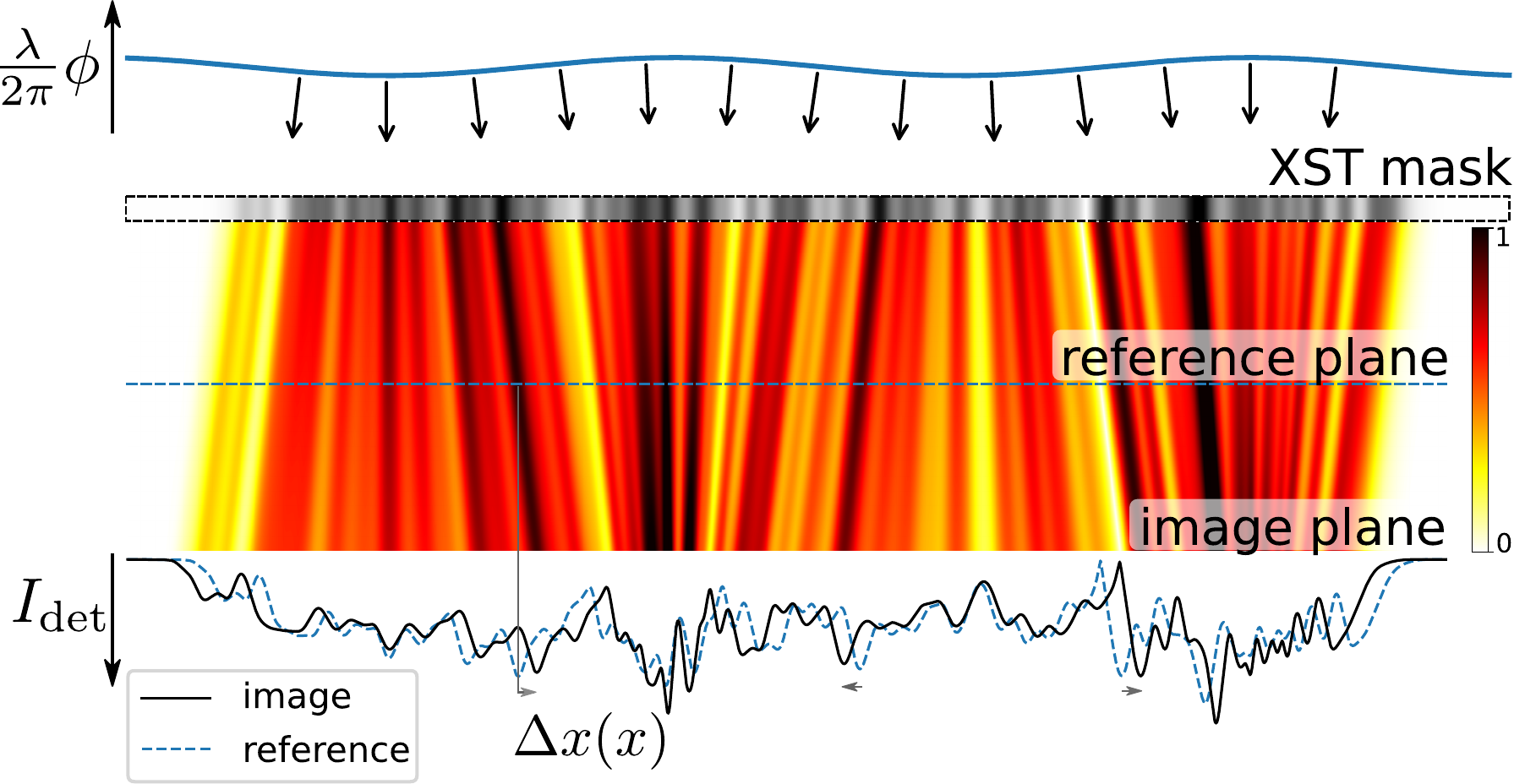}
\caption{Illustration of the X-ray Speckle Tracking (XST) principle. 
    \textbf{Top:} as in Fig. \ref{fig:Hartmann}. \textbf{Middle:} as in Fig. 
    \ref{fig:Hartmann}, with the binary mask replaced by a random phase / 
    absorption mask (dashed outline). \textbf{Bottom:} Sub-regions of the 
    measured shadow image (solid black line) are compared to the reference 
    shadow image (dashed blue line) to determine displacements (black arrows).}
\label{fig:XST}
\end{figure}

The maximum density of the holes in the grid is limited. This is because the 
task of uniquely matching each shadow image with each hole becomes more difficult 
as the hole density is increased -- a problem that is easier to appreciate in 
two dimensions. In 2012 
B\`erujon \textit{et al.} realised a simple yet elegant solution to this 
problem, one that allowed for an arbitrarily fine grid of ``masks'' with a 
resolution and sensitivity limited only by the CCD pixel array and the signal 
to noise ratio \cite{Berujon2012}.  
Their solution, the XST, is to replace the binary mask of identical holes with a 
thin random phase object, such as a diffuser, as shown in Fig. \ref{fig:XST}. 
Because the diffuser is 
random, the shadow from each sub-region of the diffuser is unique - encoded by the 
speckle pattern seen on the detector - so that one can therefore consider any 
point in the diffuser to be the centre of a virtual Hartmann hole. In this 
sense, the random object serves as a high density fiducial marker for each of 
the light rays that pass from the reference or mask plane to the detector.

In the Hartmann sensor, it is assumed that the mask is well characterised, so 
that the shadow positions can be compared to their ideal positions, which are 
known \textit{a priori}. However, since the mask is no longer a simple 
geometric object, it is now necessary to record a reference image of the mask 
with which to compare the distorted image. 

In addition to measurements of a wavefronts phase, the XST principle can be extended 
to incorporate phase imaging of arbitrary samples. This can be achieved by recording
an image of the wavefront with the diffuser (acting as a mask) in the beam path -- this
image is called the ``reference'' image. Then, another image is recorded with an 
additional sample (the one to be imaged) placed in the beam path, in addition to the 
diffuser -- this is referred to simply as the ``image''. Here the 
relative displacements between the ``reference'' and the ``image'' are due, not to 
the phase profile of the wavefront, which effects both images equally, but to the
phase profile of the sample transmission function. 

These are two XST imaging configurations suggested by B\`erujon \textit{et al.}, 
one for imaging samples and the other for wavefront metrology:

\begin{itemize}
    \item[(i)] in the \textbf{differential configuration} a speckle image is 
    recorded with and without the addition of a sample, and
\item[(ii)] in the \textbf{absolute configuration} a speckle image is recorded at 
    two detector distances with respect to the mask.
\end{itemize}

In (i), the relative motion of speckles reveals the local phase gradient of the 
sample in the beam. Whereas in (ii), the total wavefront phase is recovered and 
is therefore useful for characterising x-ray beamline optics (this is the 
configuration shown in Fig. \ref{fig:XST}). Of course, it is still possible to 
characterise beamline optics in (i), just not \textit{in-situ}, by placing the optical
element in the sample position. This approach has been useful, for 
example, in measuring the phase profile of compound refractive lens systems 
\cite{Berujon2012a} but is impractical for larger systems such as Kirkpatrick-Baez mirrors. 

Since the proposal by B\`erujon \textit{et al.} there have been a number of 
substantial improvements; see for example the extensive review by
\citeasnoun{Zdora2018}. For example, \citeasnoun{Zanette2014} developed a method 
where a diffuser is scanned so as to obtain a number of reference / image 
pairs at different diffuser positions. This step can add a great deal of 
redundancy, which improves the angular sensitivity of the method and even 
allows for multi-modal imaging of the sample when employed in the differential 
configuration. In subsequent publications, this approach has 
been termed the Unified Modulated Pattern Analysis (UMPA) method 
\cite{Zdora, Zdora:18}.

In the absolute configuration, where the reference and image have been recorded  
at two detector distances, the 
smallest resolvable angular displacement (the angular sensitivity) is given by 
the ratio of the effective pixel 
size, which is the smallest resolvable displacement of a speckle, to the 
distance between the reference and image planes $\Delta \theta = d/\Delta z$. 
Therefore, the best accuracy is obtained by maximising the distance between the 
reference and image planes. However for highly divergent wavefields, as would 
be produced (for example) by a high numerical aperture lens system, there 
arises an unavoidable trade off between the wavefront sampling frequency 
and 
the angular sensitivity. In this situation the ideal location of the 
image plane is as far downstream of the lens focus as is required to fill the 
detector array with the beam, as this maximises the wavefront sampling frequency. 
In order to minimise $\Delta \theta$ (maximise $\Delta z$) one 
should then place the reference plane as close as possible to the beam focus. 
But in this plane, the footprint of the beam on the detector may be 
much smaller than in the image plane due to the beam divergence. This leads 
to a poorly sampled reference image, as only a few pixels will span the wavefront's
footprint. Therefore, the smallest resolvable speckle shift will be larger 
than that obtainable by plane wave illumination, by a factor proportional to the 
beam divergence.  

Realising this, \citeasnoun{Berujon2014} devised an XST technique, 
X-ray Speckle Scanning (XSS), that relies on small displacements of the XST mask between acquired images. 
No reference image or images are required and the 
diffraction data is recorded in a single plane. This enables the sampling 
frequency to be maximised by placing the detector such that the divergent beam 
fills the pixel array. Without a reference image however, the speckle locations in 
one image are instead compared to the locations observed in neighbouring 
images. As the speckle displacements in each image are proportional to the 
phase gradient, the differential of the speckle locations between images are 
proportional to the second derivative of the phase; thus it can be viewed as a 
wavefront curvature measurement. The achievable angular sensitivity is now 
proportional to the step size of the mask, which can be substantially smaller 
than the effective pixel size. Interestingly, this approach is similar in 
principle to the Wigner-distribution deconvolution approach described in 
\cite{Chapman1996}. 

\begin{figure}
\includegraphics[width=\textwidth]{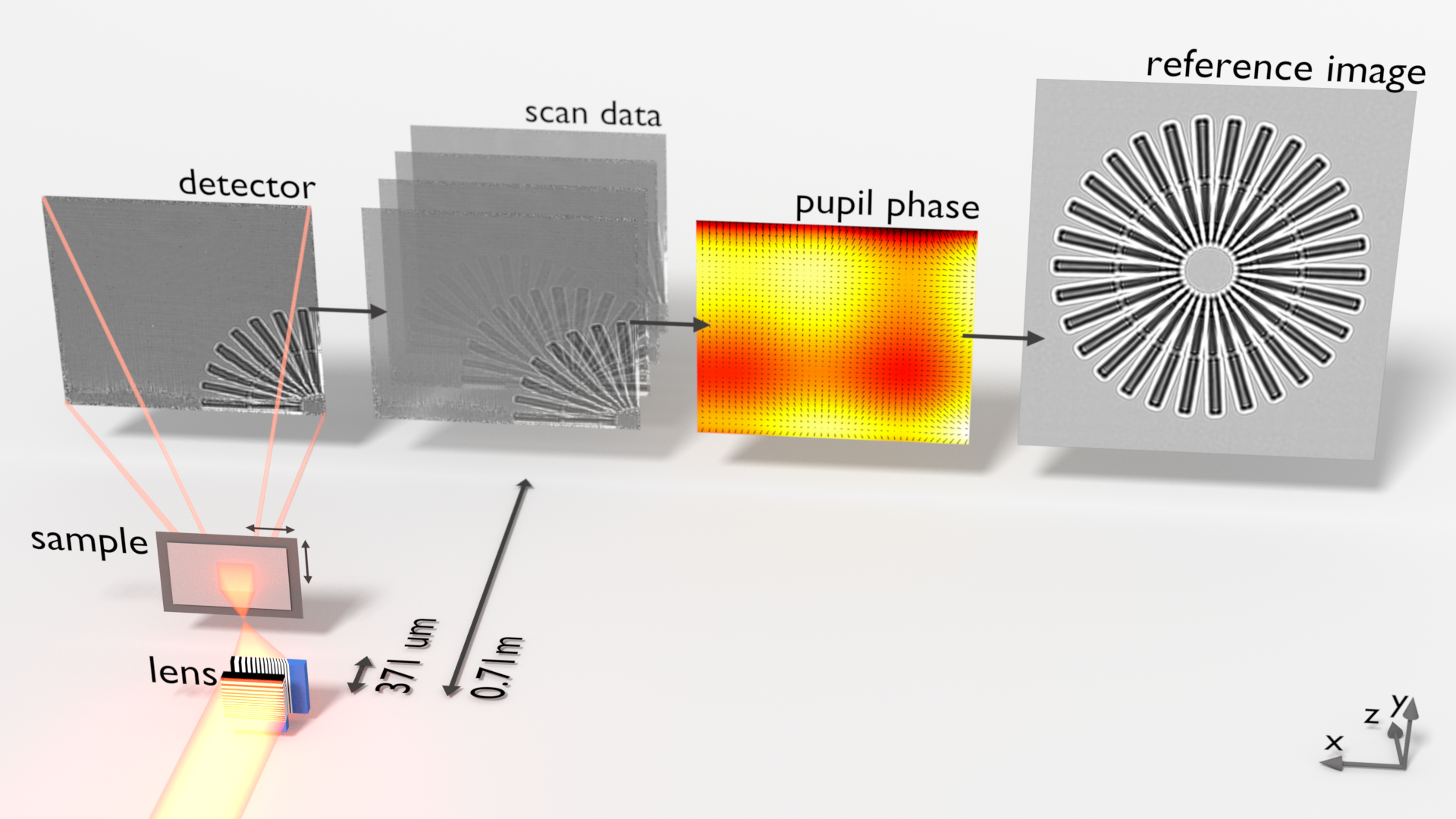}
\caption{Illustration of the ptychographic-XST method. The beamline 
    illumination was focused (off-axis) in 2D by two linear focusing lenses, with
    numerical apertures of 0.015 (horizontal) and 0.014 (vertical). 
    The Siemens star sample was placed $371\;\mu$m downstream of the 
    focal plane. Images were recorded on a CCD pixel array detector $0.71\;$m 
    downstream of the focus. The scan data consists of 49 shadow images, 
    recorded as the sample was translated across the beam profile. The wavefront phase 
    and reference image maps were refined iteratively. } 
\label{fig:overview}
\end{figure}

In the following section, we describe an approach that is similar in principle to the one described above:
\begin{itemize}
    \item[(iii)] \textbf{Ptychographic-XST:} shadow images are recorded as the mask / 
    object is translated across the wavefront. 
\end{itemize}
In this method (see Fig. \ref{fig:overview}) the unknown object acts as both 
the imaging target and the speckle mask simultaneously. There is no 
special reference image, rather each image serves as a reference for all other 
images. Both the wavefront phase (without the influence of the object) as well 
as the object image (without the influence of wavefront distortions) are determined 
in an iterative update procedure. At each iteration, 
speckles\footnote{
In this article we use the word ``speckle'' loosely, to mean any localised diffraction feature recorded by the detector.
Indeed, our method could just as well have been referred to as ``ptychographic x-ray \textit{feature} tracking''.} in the 
recorded images are compared with the current estimate of the reference image 
(in contrast to the XSS method). 
Images are recorded at a fixed 
detector distance and there is no trade off between phase sensitivity and the 
wavefront sampling frequency, making this method suitable for highly divergent 
beams. Because the speckle displacements are compared between the image and the 
estimated reference, large angular distortions can be accommodated. This is 
advantageous because it allows for the sample to be placed very near the beam 
focus, where the phase gradients across the sample surface are largest and 
where the magnification factor allows for high imaging resolution and angular 
sensitivity.

\section{The Speckle Tracking approximation}\label{sec:method}

In this section we describe the governing equation that relates the measured 
intensities in each image and the reference in terms of the wavefront phase. 
For monochromatic light, in the Fresnel diffraction regime the image formed on 
a detector placed a distance $z$ downstream of an object is given by
\begin{align}\label{eq:Fr}
I_\text{ref}(\textbf{x}, z) &= \frac{1}{(\lambda z)^2} \bigg| \iint T(\textbf{x}') 
             e^{i\pi \frac{|\textbf{x}-\textbf{x}'|^2}{\lambda z}} d\textbf{x}'\bigg|^2 ,
\end{align}
where $T(\textbf{x})$ represents the exit-surface wave of the light in the plane $z=0$. 
For plane wave illumination, under the projection approximation [see Eq. (2.39) of \cite{Paganin2006}], $T(\textbf{x})$ also 
represents the transmission function of the object. 

Now let us suppose that, rather than plane-wave illumination, the object is 
illuminated by a wavefront with an arbitrary phase ($\phi$) and amplitude ($\sqrt{w}$) profile given 
by $p(\textbf{x}, 0)=\sqrt{w}(\textbf{x})e^{i\phi(\textbf{x})}$. The observed intensity is now given by
\begin{align}\label{eq:I}
I(\textbf{x}, z) &= \frac{1}{(\lambda z)^2} \bigg| \int T(\textbf{x}') p(\textbf{x}', 0) 
           e^{i\pi \frac{|\textbf{x}-\textbf{x}'|^2}{\lambda z}} d\textbf{x}'\bigg|^2.
\end{align}
\onecolumn
\twocolumn
For XST based techniques, the challenge is to relate the image ($I$) to the 
reference ($I_\text{ref}$) via a geometric transformation. Here, the reference image 
as defined in Eq. \ref{eq:Fr} represents an image of the sample, a distance $z$ downstream of 
the sample plane, that is undistorted by
wavefront aberrations or magnified by beam divergence. For the purposes of this section, 
$I_\text{ref}$ could be a recorded image, but in following sections we will see that this
image can be estimated from a set of distorted images. 

We should note that at this 
point the mathematical description is rather general. For example, in the 
differential configuration of XST, $T(\textbf{x})$ would represent the wavefront 
generated by the diffuser in the plane of the object and $p(\textbf{x},0)$ would 
represent the transmission function of the object. In what follows however, we will 
continue to describe $T(\textbf{x})$ as the object or mask transmission function and 
$p(\textbf{x}, z)$ as the x-ray beam profile (un-modulated by the object). 

A common approach to this problem is outlined by \citeasnoun{Zanette2014}. 
There, $\phi$ is expanded up to 
first-order, and $\sqrt{w}$ to zeroth-order, in a Taylor series about the point 
$\textbf{x}$:
\begin{align}
  \phi(\textbf{x}') &= \phi(\textbf{x}) + (\textbf{x}'-\textbf{x}) \cdot \nabla \phi(\textbf{x}) + \phi_H(\textbf{x}'), \\
  \sqrt{w}(\textbf{x}') &= \sqrt{w}(\textbf{x}) + \sqrt{w}_H(\textbf{x}'),
\end{align}
where $\phi_H(\textbf{x}')$ and $\sqrt{w}_H(\textbf{x}')$ are the higher order terms in the 
expansion. Now we have, for $\phi_H$ and $\sqrt{w}_H \approx 0$,
\begin{align} \label{eq:I1}
 I(\textbf{x}, z) &\approx \frac{w(\textbf{x})}{\lambda z}  \bigg| \iint T(\textbf{x}')e^{i(\textbf{x}'-\textbf{x}) \cdot  
                 \nabla \phi(\textbf{x})} e^{i\pi \frac{|\textbf{x}-\textbf{x}'|^2}{\lambda z}} 
                 d\textbf{x}'\bigg|^2 , \nonumber \\
 &= \frac{w(\textbf{x})}{\lambda z}  \bigg| \iint T(\textbf{x}') 
    e^{\frac{i\pi}{\lambda z}|\textbf{x}-\textbf{x}' - \frac{\lambda z}{2\pi}\nabla \phi(\textbf{x})|^2 } 
    d\textbf{x}'\bigg|^2 ,\nonumber \\
 &= w(\textbf{x}) I_\text{ref}(\textbf{x} - \frac{\lambda z}{2\pi}\nabla \phi(\textbf{x}), z). 
\end{align}
This confirms the intuitive assumption that the local gradient of $\phi$ at each 
position along the sample is converted into a lateral displacement of the 
speckles observed in the reference image. Equation \ref{eq:I1} serves well in 
the limit where $\phi_H$ and $\sqrt{w}_H$ approach $0$ (i.e. for smooth 
wavefronts) and is employed in a number of XST based techniques. For example in 
the UMPA approach (see Eq. 9 in \cite{Zdora2018}) the governing equation is 
given by
\begin{align} \label{eq:umpa}
 I(\textbf{x}, z) = w(\textbf{x}) \left[\bar{I}_r + D(\textbf{x}) 
           (I_\text{ref}(\textbf{x} - \frac{\lambda z}{2\pi}\nabla \phi(\textbf{x}), z) - \bar{I}_r)\right], 
\end{align}
where $\bar{I}_r$ is the mean intensity of the reference pattern and $D(\textbf{x})$ is 
a term the authors refer to as the ``dark field signal''. This term is related 
to a reduction in fringe visibility due to fine features in $w(\textbf{x})$ and, in fact, 
serves as an alternative contrast mechanism when solved for in addition to 
the phase gradients. Putting this term aside by setting $D=1$, one can see 
that Eq. \ref{eq:umpa} reduces to Eq. \ref{eq:I1}. 

Given the restrictive nature of the approximations employed however, it is not 
surprising that Eq. \ref{eq:I1} quickly fails to serve as a valid approximation 
for larger phase gradients. To see this, let us consider a well known analytic 
solution to $I$ in terms of $I_\text{ref}$ called the ``Fresnel scaling theorem'', which
is described in,
for example, appendix B of \cite{Paganin2006}. Simply put, it 
states that:
\vskip 0.1cm
\noindent \textit{The projected image of a thin scattering object from a point
source of monochromatic light is equivalent to a magnified
defocused image of the object illuminated by a point source of
light infinitely far away.}
\vskip 0.1cm
The derivation is rather simple and so we shall present it here using the 
current notation. Let us say that the image, $I$, is formed by the point source 
of illumination a distance $z_1$ along the optical axis (the $z$-axis) and that 
this distance is large enough that we can ignore intensity variations of the 
illumination across the sample surface, so that $\sqrt{w}(\textbf{x})=1$. The 
probing illumination in the plane of the sample is then 
given by $p(\textbf{x}, 0) = e^{i\pi \textbf{x}^2 / \lambda z_1}$. Substituting this into Eq. 
\ref{eq:I} and completing the square in the exponent, we have
\begin{align}\label{IF}
 I(\textbf{x}, z) &= \frac{1}{(\lambda z)^2} \bigg| \iint T(\textbf{x}') e^{i\pi \textbf{x}'^2 / \lambda z_1} 
            e^{i\pi \frac{|\textbf{x}-\textbf{x}'|^2}{\lambda z}} d\textbf{x}'\bigg|^2, \nonumber \\ 
         &= \frac{1}{(\lambda z)^2} \bigg| \iint T(\textbf{x}') 
            e^{\frac{i\pi}{\lambda} \frac{z_1 + z}{z z_1} 
            |\frac{z_1}{z_1 + z}\textbf{x} - \textbf{x}'|^2} d\textbf{x}'\bigg|^2, \nonumber \\
        &= \left(\frac{z_1}{z_1 + z}\right)^2 I_\text{ref}(\frac{z_1}{z_1 + z}\textbf{x}, 
           \frac{z z_1}{z_1 + z}) \nonumber \\
        &= M^{-2} I_\text{ref}(\textbf{x} /M, z/M),
\end{align}
where the geometric magnification factor $M = \frac{z_1 + z}{z_1}$ and $z/M$ is the effective propagation distance ($\bar{z}$). But 
according to Eq. \ref{eq:I1} we would have
\begin{align}\label{I1p}
I(\textbf{x}, z) &= I_\text{ref}(\textbf{x} - \frac{\lambda z}{2\pi} \frac{2\pi \textbf{x}}{\lambda z_1}, z) \\
        &= I_\text{ref}(\frac{z_1-z}{z_1} \textbf{x}, z),
\end{align}
with a geometric magnification factor $M' = \frac{z_1}{z_1 - z}$, in 
contradiction to the result from the Fresnel scaling theorem. As expected, the 
results agree in the limit $z_1\rightarrow \infty$, i.e. in the limit where the 
phase gradient approaches $0$. Current formulations for XST based on Eq. 
\ref{eq:I1} (in the absolute configuration), are expected to perform badly when 
the effective source distance, $z_1$, approaches the propagation distance, $z$, 
or, in the differential configuration, when the sample transmission function 
departs significantly from the weak phase approximation. 

In a notable departure from this approach, Pagenin \textit{et al.} have 
recently developed an alternative description of the speckle tracking 
approximation based on a ``geometric flow'' equation \cite{Paganin2018}
\begin{align}\label{eq:pag}
 I(\textbf{x}, z) &\approx I_\text{ref}(\textbf{x}, z) - \frac{\lambda z}{2\pi} \nabla \cdot 
                  \left( I_\text{ref}(\textbf{x}, z)  \nabla \phi(\textbf{x}) \right). 
\end{align}
This approximation, which closely resembles the transport of intensity equation \cite{Teague1983}, has the remarkable property that $\phi$ may be determined 
analytically from a reference-image pair, thus permitting the rapid and 
simple processing of large tomographic data sets. This approach also 
assumes small and local distortions of the reference image and is, therefore, 
ill-suited as an approximation for larger phase gradients. For example, 
substituting the quadratic phase for a diverging wavefield, $\phi = \frac{\pi \textbf{x}^2}{\lambda z_1}$, into Eq. \ref{eq:pag} yields
\begin{align}\label{eq:Pag}
 I(\textbf{x}, z) &\approx \frac{z_1-z}{z_1} \left[ I_\text{ref}(\textbf{x}, z) - 
                  \frac{z}{z_1-z} \textbf{x} \cdot \nabla I_\text{ref}(x, z) \right]
\end{align}
This corresponds to a geometric magnification factor of $M'' = 
\frac{z_1 - z}{z_1-2z}$, once again, in contradiction to the analytic result 
$M=\frac{z_1 + z}{z_1}$. 

To see this more clearly, let us examine the exact result of Eq. \ref{IF} in the limit where 
$M\rightarrow 1$. First, we set $1/M = 1 + m$, so that $m\rightarrow 0$ as 
$M\rightarrow 1$. Then we expand $I_\text{ref}(\textbf{x}/M, z/M)$ to first-order in a Taylor 
series about $\textbf{x}$: 
\begin{align*}
I(\textbf{x}, z) &= M^{-2} I_\text{ref}(\textbf{x} + m\textbf{x}, z/M) \\
        &\approx M^{-2} \left[ I_\text{ref}(\textbf{x}, z/M) + m \textbf{x} \cdot \nabla I_\text{ref}(\textbf{x}, z/M)\right].
\end{align*}
Comparing the above equation with Eq. \ref{eq:Pag}, we have $m = 
\frac{-z}{z_1-z}$. Solving for the geometric magnification factor yields 
$M'' = \frac{z_1 - z}{z_1-2z}$ as above. 

\begin{figure}
\includegraphics[width=8.88cm]{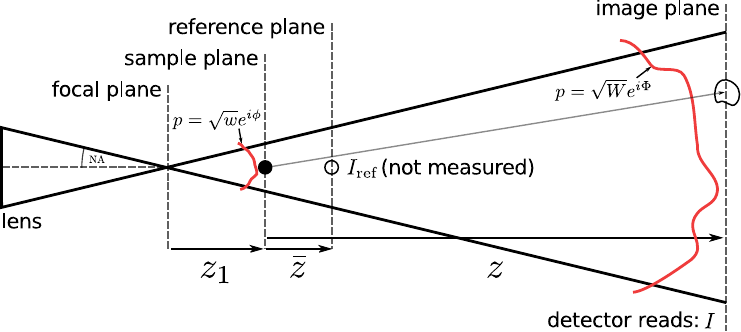}
\caption{Schematic diagram for a hypothetical projection imaging experiment. The 
illuminating beam propagates from left to right and the solid black lines 
indicate the boundaries of the illumination wavefront. The sample is depicted 
as a small black filled circle in the sample plane and as a black circle in the reference and image planes. The red lines depict the illumination's wavefront in the sample and image planes, which are not merely related by transverse magnification. The distorted shape of the circle in the image plane represents possible distortions of the speckle produced by the sample and the transverse phase gradients of the illumination.}
\label{fig:stem}
\end{figure}

Remarkably, with only a minor modification to the speckle tracking formula in 
Eq. \ref{eq:I1}, a second-order expansion of the phase term can be accommodated 
in the Fresnel integral, leading to
\begin{align} 
    I(\textbf{x}, z) &\approx W(\textbf{x}) I_\text{ref}(\textbf{x} - \frac{\lambda z}{2\pi} \nabla \Phi(\textbf{x}), 
                     \bar{z}), \label{eq:Fr72} \\
    \left(\frac{\bar{z}}{z}\right)^2 w(\textbf{x}) I_\text{ref}(\textbf{x}, \bar{z}) &\approx  I(\textbf{x} + \frac{\lambda z}{2\pi} \nabla \phi(\textbf{x}), z), \label{eq:Fr82}
\end{align}
%
where $\nabla \phi$ and $\nabla \Phi$ are the transverse gradients of the illuminating wavefield phase in the sample and image  planes respectively (without the influence of the object) and $w$ and $W$ are the intensity profiles of the illuminating wavefield in the reference and image planes respectively. In Fig. \ref{fig:stem} we show a diagram for a hypothetical PXST imaging experiment. In this diagram one can see the lens, focal, sample, reference and image planes respectively. The reference image would have been measured by plane-wave illumination in the plane indicated. A point that is not illustrated in the diagram, is that both the image and the reference image exhibit propagation effects, such as Fresnel fringes. We note, once again, that the speckle tracking approximation above, applies to more imaging geometries / modalities than that displayed in Fig. \ref{fig:stem}.

Equations \ref{eq:Fr72} and \ref{eq:Fr82} are reciprocal statements of the same approximation and choosing between them is a matter of convenience depending on the desired application. We note here that this approximation makes a distinction between the phase gradients in the sample and image planes, whereas it is common to assume that they are similar or related by a lateral scaling factor (magnification). This distinction is not important in cases where the separation between these two planes and the beam divergence are small, but becomes critical for highly magnified imaging geometries or long propagation distances. This approximation is not as strong as the ``stationary phase approximation'' \cite{Fedoryuk1971}, which links coherent propagation theory with geometric optics, although the principles used to derive this result are similar. The derivation is straightforward and self-contained but lengthy, and may be found in section \ref{sec:approx} of the appendix.

Equations \ref{eq:Fr72} and \ref{eq:Fr82} posses two beneficial properties for the current analysis: they relate the image and its reference via a geometric transformation and they are consistent with the Fresnel scaling theorem. In fact, the Fresnel scaling theorem is a special case of the above approximations when $w(\textbf{x}) = 1$ and $\phi(\textbf{x}) = \pi \frac{\textbf{x}^2}{\lambda z_1}$. Evaluating Eq. \ref{eq:Fr82} for these values of $w$, $\phi$ and using $\bar{z} = z z_1 / (z+z_1)$ we have
\begin{align*} 
 I_\text{ref}(\textbf{x}, \frac{z z_1}{z+z_1}) &= \left(\frac{z_1+z}{z_1}\right)^2 I(\textbf{x} + \frac{\lambda z}{2\pi} \frac{2\pi \textbf{x}}{\lambda z_1}, z),\\
 &= \left(\frac{z_1+z}{z_1}\right)^2 I(\textbf{x} \frac{z_1 + z}{z_1}, z),
 \\
  \text{and so } I(\textbf{x}, z) &=  \left(\frac{z_1+z}{z_1}\right)^{-2} I_\text{ref}(\textbf{x} \frac{z_1}{z_1 + z}, \frac{z z_1}{z_1+z}),
\end{align*}
which yield the correct magnification and scaling factors, in agreement with Eq. \ref{IF}. Similarly, we can evaluate Eq. \ref{eq:Fr72} using
\begin{align}
 W(\textbf{x})    &= \left(\frac{\bar{z}}{z}\right)^2, &
 \Phi(\textbf{x}) &= \frac{\pi \textbf{x}^2}{z_1 + z},
\end{align}
where these values for the illumination's wavefront in the plane of the detector follow from the Fresnel approximation for a point source placed a distance $z+z_1$ upstream and from flux conservation of the beam when $w(\textbf{x})=1$ in the sample plane. 

Evaluating Eq. \ref{eq:Fr72} yields
\begin{align}
 I(\textbf{x}, z) &= \left(\frac{z_1}{z+z_1}\right)^2 I_\text{ref}(\textbf{x} - \frac{\lambda z}{2\pi} 
            \frac{2\pi \textbf{x}}{z_1 + z} , \frac{z z_1}{z+z_1}),  \\
         &= \left(\frac{z_1}{z+z_1}\right)^2 I_\text{ref}(\textbf{x}\frac{z_1}{z_1 + z} , \frac{z z_1}{z+z_1}),
\end{align}
which is once again, in agreement with Eq. \ref{IF}.

In general, for arbitrary $\phi$, the phase curvature of the illumination may vary in direction, as is the case (for example) in an astigmatic lens system, and also with position in the image. Thus, the magnification is also position dependent and directional: 
\begin{align}
 M_\textbf{v}(\textbf{x}) &= \left[1 - \frac{\lambda z}{2\pi} \nabla^2_\textbf{v} \Phi(\textbf{x}) \right]^{-1},
\end{align}
where $\nabla_\textbf{v} \Phi(\textbf{x})$ is the directional derivative of $\Phi(\textbf{x})$ along the unit normal vector $\textbf{v}$.

Given the extended validity of Eq. \ref{eq:Fr72}, we suggest that the following modification to the UMPA equation (Eq. \ref{eq:umpa}), will achieve better results:
\begin{align} 
    I(\textbf{x} + \frac{\lambda z}{2\pi} \nabla \phi(\textbf{x}), z) 
    &\approx w(\textbf{x}) \left[\bar{I}_r + D(\textbf{x})(I_\text{ref}(\textbf{x}, z) - \bar{I}_r) \right],
\end{align}
where $\bar{I}_\text{ref} = \langle I_\text{ref}(\textbf{x}) \rangle_\textbf{x}$, or, using the notation of \citeasnoun{Zdora2018}:
\begin{align} 
    I(x - u_x, y - u_y) 
    &\approx T(x, y) \left[\bar{I}_0 + D(x, y)(I_0(x, y) - \bar{I}_0) \right].
\end{align}

We also note that although Paganin \textit{et al.}'s geometric flow algorithm (Eq. \ref{eq:pag}) is a poor approximation for larger distortion factors (large $M$), it may be a more general physical description in the limit $M\rightarrow 1$. As the authors note, the term $\propto \nabla I_\text{ref} \cdot \nabla \phi$ in the expansion of Eq. \ref{eq:pag} accounts for speckle translations that arise from strong intensity gradients of the reference image, i.e. that are not generated from $\nabla \phi$ alone.

\section{Limits to the Approximation}\label{sec:limits}

The second-order speckle tracking approximation of Eqs \ref{eq:Fr72} and 
\ref{eq:Fr82} are subject to the following approximations
\begin{align*}
 &\text{1:} &g(\textbf{x}, \textbf{x}') &\approx g(\textbf{x}, \textbf{u}(\textbf{x})) + 
             \frac{1}{2} \bigg[ \nabla^2\phi(\textbf{u}(\textbf{x})) + 
             \frac{2\pi}{\lambda z}\bigg] |\textbf{x}'-\textbf{u}(\textbf{x})|^2, \\
 &\text{2:} &\sqrt{w}(\textbf{x}') &\approx \sqrt{w}(\textbf{u}(\textbf{x})), \\
 &\text{3:} &z(\textbf{x}) &\approx \bar{z} \equiv \frac{z z_1}{z+z_1},  
\end{align*}
where these are additional to the approximations necessary for the paraxial 
approximation to hold, $\textbf{u}(\textbf{x}) = \textbf{x} - \frac{\lambda z}{2\pi} \nabla \Phi(\textbf{x})$ and 
$g(\textbf{x}, \textbf{x}') \equiv \pi\frac{|\textbf{x}-\textbf{x}'|^2}{\lambda z} + \phi(\textbf{x}')$. In general, these approximations hold best for smooth wavefront amplitudes $\sqrt{w}$, predominantly quadratic phase $\phi$ and large spatial frequencies of the object. 

Here, we examine the speckle tracking approximation, in 1D, for the imaging geometry depicted in Fig.
\ref{fig:stem} and with parameters corresponding to a typical experiment utilising x-ray multilayer Laue lenses. 
For this example we choose that the illumination is formed by a lens with a hard edged 
aperture and with the sample placed at two possible distances from the focal plane, $z_1=500\;\mu$m and $z_1=10\;\mu$m. The lens has a numerical aperture of $\text{NA}=0.01$ and the detector is placed in the far-field of the probe and 
the sample, with $z_1 + z = 1\;$m. This imaging geometry leads to an effective 
propagation for plane-wave illumination that is nearly identical to the distance from the focus to the sample ($\bar{z}\approx z_1$). 
The wavelength is
$10^{-9}$m. The sample has a Gaussian profile so 
that $T(x) = 1 - n e^{-x^2/2\sigma^2}$, where $n=1-i$ was chosen arbitrarily 
and would be proportional to the sample thickness and the deviation from unity of the 
refractive index. The Fresnel number is thus $F\approx \sigma^2 / \lambda \bar
{z}$.  

\onecolumn
\begin{figure}
\includegraphics[width=\textwidth]{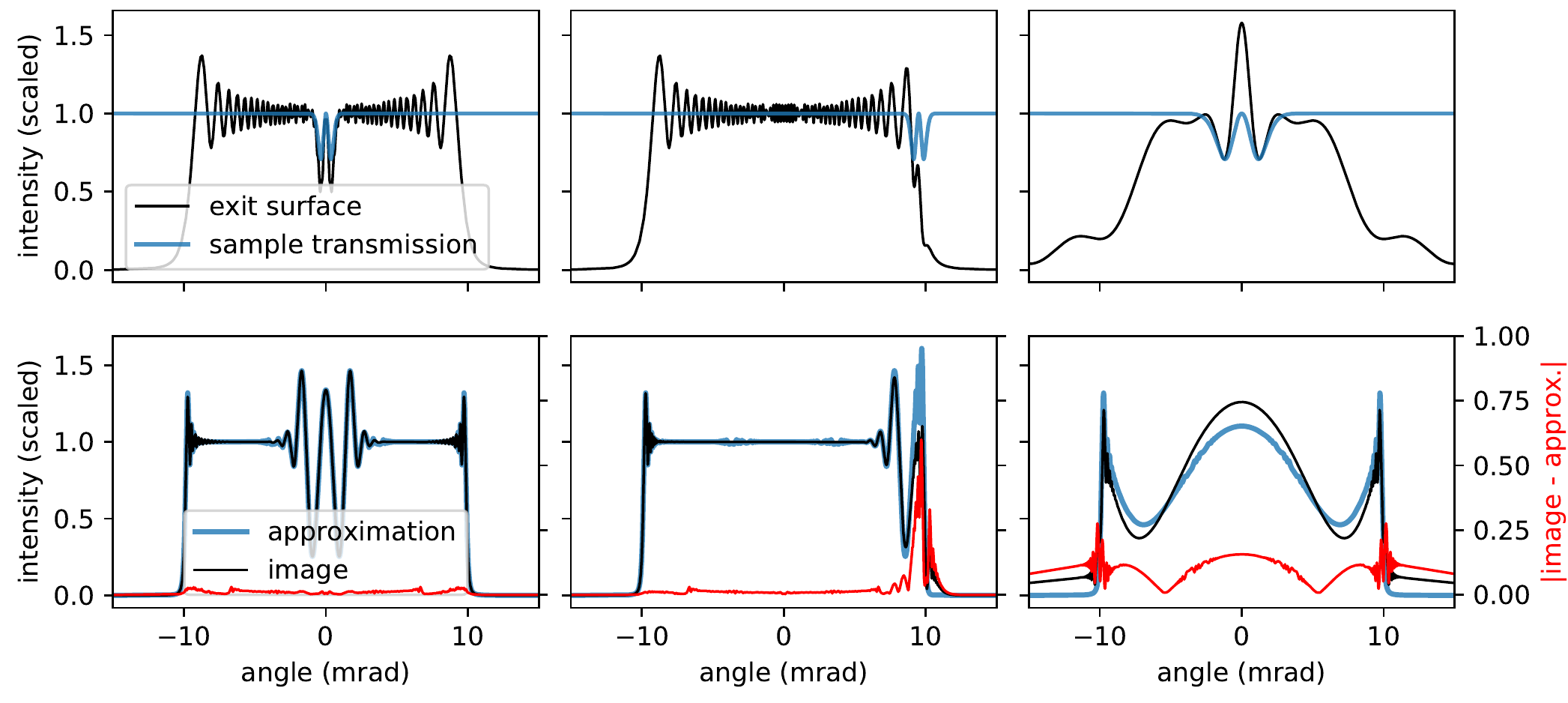}
\caption{Comparison between the images formed according to Frensel diffraction 
theory and the speckle tracking approximation. In the left and middle columns, the sample has a $\sigma$-width of $0.15\mu$m and is placed $500\mu$m from the focus. In the left column the sample is centred in the beam profile, whilst in the middle it has been shifted to the edge. In the right column the sample has a $\sigma$-width of $0.01\mu$m and is placed $10\mu$m from the focus.
\textbf{First row}: the exit surface wave intensities formed by illuminating a 
small Gaussian object with divergent illumination (black line). The 
intensities have been scaled by the factor $z/\bar{z}$. The sample 
transmission amplitudes are shown in blue. The angles along the x-axis are given by $\arctan{x/z_1}$ and match those of the second row. \textbf{Second row}: the intensity of the wavefront in the image plane 
(black line) and the images formed by the speckle tracking approximation 
(blue line). The fractional differences are shown in red. The angles are given 
by $\arctan{x/(z_1 + z)}$.}
\label{fig:stem_im}
\end{figure}
\twocolumn

The wavefronts in the sample and image planes were simulated using the 
discrete form of the Fresnel diffraction integral. The illumination's 
wavefront in the image plane is given by $p(x, z) = c \sqrt{W}(x) e^{i \Phi(x)}$, where c is a complex pre-factor that does not depend on $x$, $\sqrt{W}(x)$ was calculated numerically and $\Phi$ is almost quadratic, with $\Phi(x) \approx \frac{\pi x^2}{\lambda (z_1 + z)}$. Note that $\phi(x)$, the phase profile of the illumination in the sample 
plane, is not given by $\frac{\pi x^2}{\lambda z_1}$ as would be the 
case for a point source of light (i.e. for $\text{NA}\rightarrow \infty$). 
This is because the hard edges of the aperture produce Fresnel fringes that 
progress from the edge of the wavefront to the focal point at $x=0$ as one 
moves from the image to the focal plane.

To test the validity of the speckle tracking approximation, we compare these 
simulated Fresnel images with those formed by evaluating Eq. \ref{eq:Fr72}. In 
this case Eq. \ref{eq:Fr72} can be evaluated analytically with
\begin{align} \label{eq:approx}
 W(x) I_\text{ref}(x-\frac{\lambda z}{2\pi}\nabla \Phi(x), \bar{z}) 
 &= W(x) \bigg| 1 + n \frac{\sigma}{\sigma'} 
    e^{-\frac{x^2}{2M^2 \sigma'^2}} \bigg|^2 ,\\
 \text{where } \sigma'^2 &= \sigma^2 + i \frac{\lambda \bar{z}}{2\pi}, \\
 \text{and }  x-\frac{\lambda z}{2\pi}\nabla \Phi(x) 
 &= x \frac{z_1}{z_1 + z} = \frac{x}{M}.
\end{align}
In order to arrive at the above result, we have assumed that $\Phi$ is purely quadratic across the wavefront, 
but this approximation has not been used when simulating the image according to Fresnel diffraction theory. 

In appendix \ref{sec:limits_der}, we suggest a suitable criterion for the speckle tracking approximation to hold for this imaging geometry based on the second criterion above,
\begin{align} \label{eq:dwdq_stem2}
 \frac{\sqrt{\lambda \bar{z}} + \lambda \bar{z} q_T}{z_1 \text{NA}} &\ll 1,
\end{align}
where $q_T=1/X$ is the spatial frequency corresponding to full period features of size $X$. This criterion holds for features within the plateau of the illumination profile. 

In the first column of Fig. \ref{fig:stem_im}, we have placed the sample in 
the centre of the illumination profile. Here, the left hand side of Eq. \ref{eq:dwdq_stem2} evaluates to $0.8$ and one can see that the fractional differences between the image and the approximation are small compared to that of the middle column. There, the sample has been shifted to the edge of the illumination profile, where the slope of the illumination amplitude is large. This leads to a breakdown of the second condition ($\sqrt{w}(x') \approx \sqrt{w}(u(x))$) and indeed the discrepancy between the approximation and the image is largest near the edge of the pupil region and slowly reduces for features closer towards the central region.  

In the right column of Fig. \ref{fig:stem_im}, the sample is smaller, with a $\sigma$-value of $0.01\mu$m and has been moved closer to the focal point. The left hand side of Eq. \ref{eq:dwdq_stem2} now evaluates to $11.0$. As expected, this increase from $0.8$ in the first column to $11.0$ in the right column corresponds to an increasing discrepancy between the speckle tracking approximation and the image. This image is in the transition region between the near-field and far-field diffraction regimes. Clearly, features in the diffraction outside of the holographic region, where $\sqrt{W} \approx 0$, are not represented at all by the approximation.

In both the second and third examples shown here, the errors in the speckle tracking  
approximation are dominated by the error in the approximation $\sqrt{w}(x') 
\approx \sqrt{w}(u(x))$. This is not surprising given that the zeroth-order 
expansion of $\sqrt{w}(x')$ about $u(x)$ is a much stronger approximation than 
the second-order expansion of $\phi(x')$ about $u(x)$ (both approximations 
are necessary to arrive at the speckle tracking formula). 

The increased quality of projection images due to smoother illumination 
profiles was one of the principle motivations behind Salditt and collaborators' 
efforts to develop an x-ray single-mode waveguide, in order to improve 
their tomo-holographic imaging methods; see for example \cite{Krenkel2017}.

\section{Reconstruction Algorithm}\label{sec:algorithm}


In this section we describe the steps necessary to recover estimates for $\Phi(\textbf{x})
$ and $I_\text{ref}(\textbf{x})$ from a series of $N$ measurements of the kind depicted in Fig. 
\ref{fig:overview}; where each recorded image on the detector corresponds to a 
translation of the sample in the transverse plane by $\Delta \textbf{x}_n$ (here $n$ is 
the image index). According to the speckle tracking approximation of Eq. \ref
{eq:Fr72}, the geometric relationship between the recorded images $I_n(\textbf{x})$ and 
the \textit{un}-recorded reference image $I_\text{ref}(\textbf{x})$ is given by
\begin{align}\label{eq:st}
 I_n(\textbf{x}) &= W(\textbf{x}) I_\text{ref}(\textbf{x} - \frac{\lambda z}{2\pi} \nabla \Phi(\textbf{x}) 
                      - \Delta \textbf{x}_n, \bar{z}).
\end{align}
Translating the sample by $\Delta \textbf{x}_n$ along the x-axis leads to a 
corresponding translation of the reference image, this is because the 
convolution integral in Eq. \ref{eq:Fr} possesses translational equivariance.
\begin{figure}
\includegraphics[width=8.88cm]{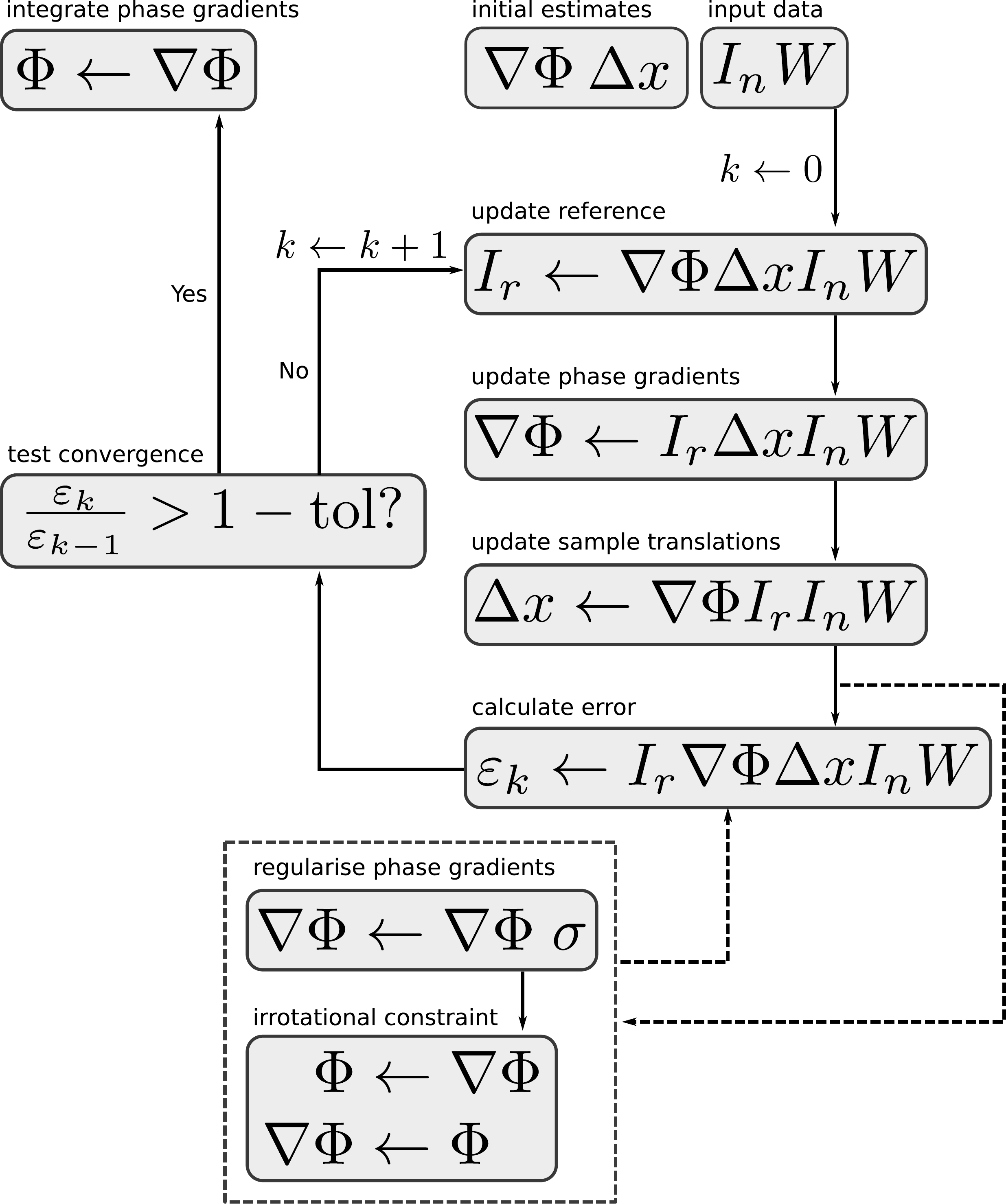}
\caption{Flow diagram for the PXST iterative refinement cycle. The 
$\leftarrow$'s represent an update of the item on the left given the items 
to the right of the arrow. 
The dashed line arrows represent optional paths in the algorithm. Each step in 
the diagram corresponds to an equation in the main text: ``update reference'' to 
Eq. \ref{eq:update_Ir}, ``update phase gradients'' to Eq. \ref{eq:update_Phi}, 
``update sample translation'' to Eq. \ref{eq:update_Dx}, ``calculate error'' to 
Eq. \ref{eq:calc_err}, ``regularise phase gradients'' to Eq. \ref{eq:reg_phase} 
and finally the ``irrotational constraint'' and ``integrate phase gradients'' 
steps are given by Eq. \ref{eq:integrate}.}
\label{fig:alg}
\end{figure}
To recover estimates for $\Phi(\textbf{x})$ and $I_\text{ref}(\textbf{x})$, we choose to minimise the 
target function
\begin{align}\label{eq:calc_err}
 \varepsilon &= \sum_n \int \varepsilon(n, \textbf{x}) dx, \\
 &= \sum_n \iint d\textbf{x} \times \nonumber \\
    & \qquad \frac{1}{\sigma^2_I(\textbf{x})} \left(I_n(\textbf{x}) -  W(\textbf{x}) 
    I_\text{ref}(\textbf{x} - \frac{\lambda z}{2\pi} \nabla \Phi(\textbf{x}) - \Delta \textbf{x}_n, \bar{z})
    \right)^2 ,\nonumber
\end{align}
in an iterative update procedure with respect to $\nabla \Phi(\textbf{x})$ and (as needed) $\Delta \textbf{x}_n$, subject to
\begin{align}\label{eq:update_Ir}
 I_\text{ref}(\textbf{x}, \bar{z}) &= \frac{\sum_n w(\textbf{x}+\Delta \textbf{x}_n) I_n(\textbf{x} + \frac{\lambda z}{2\pi} 
                    \nabla \phi(\textbf{x}+\Delta \textbf{x}_n)+\Delta \textbf{x}_n, z)}{\sum_n w^2(\textbf{x}+\Delta \textbf{x}_n) },
\end{align}
where $\sigma^2_I(\textbf{x})$ is the variance of the recorded intensities at each 
detector pixel, such that $\sigma^2_I(\textbf{x}) = \langle I^2_n(\textbf{x}) - \langle I_n(\textbf{x})
\rangle_n^2 \rangle_n$. In fact Eq. \ref{eq:update_Ir} is the analytic solution 
for the minimum of $\sum_n \varepsilon(n, \textbf{x})$ with respect to $I_\text{ref}(\textbf{x})$ but for $
\sigma^2_I(\textbf{x}) = 1$. The reason we have set $\sigma^2_I(\textbf{x}) = 1$ for the 
reference image update is that, in this way, the reference image is formed 
preferentially from parts of the image with larger intensities and thus will not 
be unduly effected by detector noise. This is also the update procedure that is 
often employed in single-mode ptychographic reconstructions. 

The update for $\nabla \Phi(\textbf{x})$ is given by
\begin{align}\label{eq:update_Phi}
 \nabla \Phi(\textbf{x}) &= \text{argmin}_{\nabla \Phi} \left[ \sum_n \varepsilon(n, x) \right],
\end{align}
while holding $I_\text{ref}(\textbf{x})$ and $\Delta \textbf{x}_n$ constant, where $\text{argmin}_{\nabla \Phi}
$ means ``the argument of the minimum'' with respect to $\nabla \Phi$, which is to 
say, the $\nabla \Phi$ that gives rise to the minimum of $\sum_n \varepsilon(n, x)$. The 
minimisation is performed by evaluating $\sum_n \varepsilon(n, \textbf{x})$ for 
possible value of $\nabla \Phi(\textbf{x})$ within a pre-defined search window. 

The update for $\Delta \textbf{x}_n$ is given by
\begin{align}\label{eq:update_Dx}
 \Delta \textbf{x}_n &= \text{argmin}_{\Delta \textbf{x}} \left[ \int \varepsilon(n, \textbf{x}) dx\right],
\end{align}
while holding $I_\text{ref}(\textbf{x})$ and $\nabla \Phi(\textbf{x})$ constant. Once again, the 
minimisation is performed by evaluating possible value of $\Delta \textbf{x}_n$ 
within a pre-defined search window. 

Additionally, it is often desirable to regularise $\nabla \Phi(\textbf{x})$ during the 
update procedure (especially for the first few iterations), according to
\begin{align}\label{eq:reg_phase}
 \nabla \Phi(\textbf{x}) &= \left( \frac{1}{2\pi \sigma^2} e^{-\frac{\textbf{x}^2}{2 \sigma^2}} \right) \otimes \nabla \Phi(\textbf{x})
\end{align}
where $\otimes$ is the convolution operator and $\sigma$ is the regularisation 
parameter that can be reduced as the iterations proceed. 

Once the iterative procedure has converged, the phase profile of the 
illumination ($\Phi(\textbf{x})$) can be recovered from the gradients ($\nabla \Phi(\textbf{x})$) 
by numerical integration. For this we follow the method outlined in the supplementary section 
of \citeasnoun{Zanette2014}. Let us label 
the final value of the phase gradients by $\delta(\textbf{x}) \equiv \nabla \Phi(\textbf{x})$.
The procedure is then given by
\begin{align}\label{eq:integrate}
 \Phi(\textbf{x}) &= \text{argmin}_\Phi\left[ \left| \delta(\textbf{x}) - \nabla \Phi(\textbf{x}) \right|^2 
            \right],
\end{align}
where $\nabla \Phi(\textbf{x})$ is evaluated numerically and the minimisation is 
performed via the least squares conjugate gradient method.

The fact that $\Phi$ is given by the numerical integration of $\nabla \Phi$ 
suggests a further constraint that could be employed in the update procedure. 
As noted by \citeasnoun{Paganin2018}: since $\nabla \Phi$ 
is given by the gradient of a scalar field, then $\nabla \Phi$ will be 
irrotational if $\Phi$ is continuous and single valued. This follows from 
the Helmholz theorem, which states that any field can be written as the sum
of a gradient and a curl. Since we know that $\nabla \Phi$ is, by definition, 
the gradient of $\Phi$, then the curl must be zero $\nabla \times \nabla \Phi = 0$.
In their work, this 
condition is automatically satisfied by the solution. Here however, we must 
incorporate this as a separate constraint. An irrotational field $\mathbf{f}$ is 
one that satisfies
\begin{align}\label{eq:irrotational}
 \frac{\partial \mathbf{f}_y(\mathbf{x})}{\partial x} &=  
                     \frac{\partial \mathbf{f}_x(\mathbf{x})}{\partial y},
\end{align}
where $\mathbf{f}_x(\mathbf{x})$ and $\mathbf{f}_y(\mathbf{x})$ are the $x$ and 
$y$ components of the vector field respectively. To ensure that $\nabla \Phi$ is 
irrotational, one need only apply the numerical integration in Eq. \ref
{eq:integrate} followed by numerical differentiation as needed during the update 
procedure. If this condition is not enforced, then the degree to which the 
recovered $\nabla \Phi$ is irrotational can be used as a measure of the fidelity 
of the result. 

Numerical considerations for the implementation of this iterative update procedure,
in addition to the source code developed to implement the PXST algorithm has been published
online\footnote{See also https://github.com/andyofmelbourne/speckle-tracking}.

The algorithm presented here is by no means the only approach to solve for the 
phase gradients and reference image. Indeed, similar problems emerge in many 
areas of imaging such as computer vision, medical imaging and military targeting
applications. In magnetic resonance imaging, the process of identifying the 
distortions that relate an image to its reference is often termed the 
``image registration'' problem and generating the reference image from a set of 
distorted views is termed ``atlas construction''. So called ``diffeomorphic image
registration'' algorithms are popular in that field, many of which are based on 
Thirion's demons and log-demons algorithm \cite{Thirion1998, Lombaert2014}. 
Indeed, this approach has been employed in the context of XST by Guillon \textit{et al.} 
\cite{Berto:17} to recover the phase gradients from an image/reference pair. 
Others in the XST field use correlation based approaches, where the geometric
 mapping between 
a small region of the distorted image and the reference is determined by the point
which provides the greatest correlation correlation coefficient \cite{Zdora2018}. 
The approach outlined in this work was employed because of its simplicity and ease 
of implementation. However, it seems likely (in the authors' view) that one or 
more of these approaches could be adapted to the current problem in order to 
produce superior results.


\subsection{Example reconstruction}

Here we provide a brief example of a PXST reconstruction from a simulated 1D dataset. This example is not 
intended as realistic simulation for an actual experiment, see \cite{Morgan2019a} for experimental results in 2D. Rather, it serves as a simple illustrative check on the basic principles of PXST. 

\begin{figure}
\includegraphics[width=8.88cm]{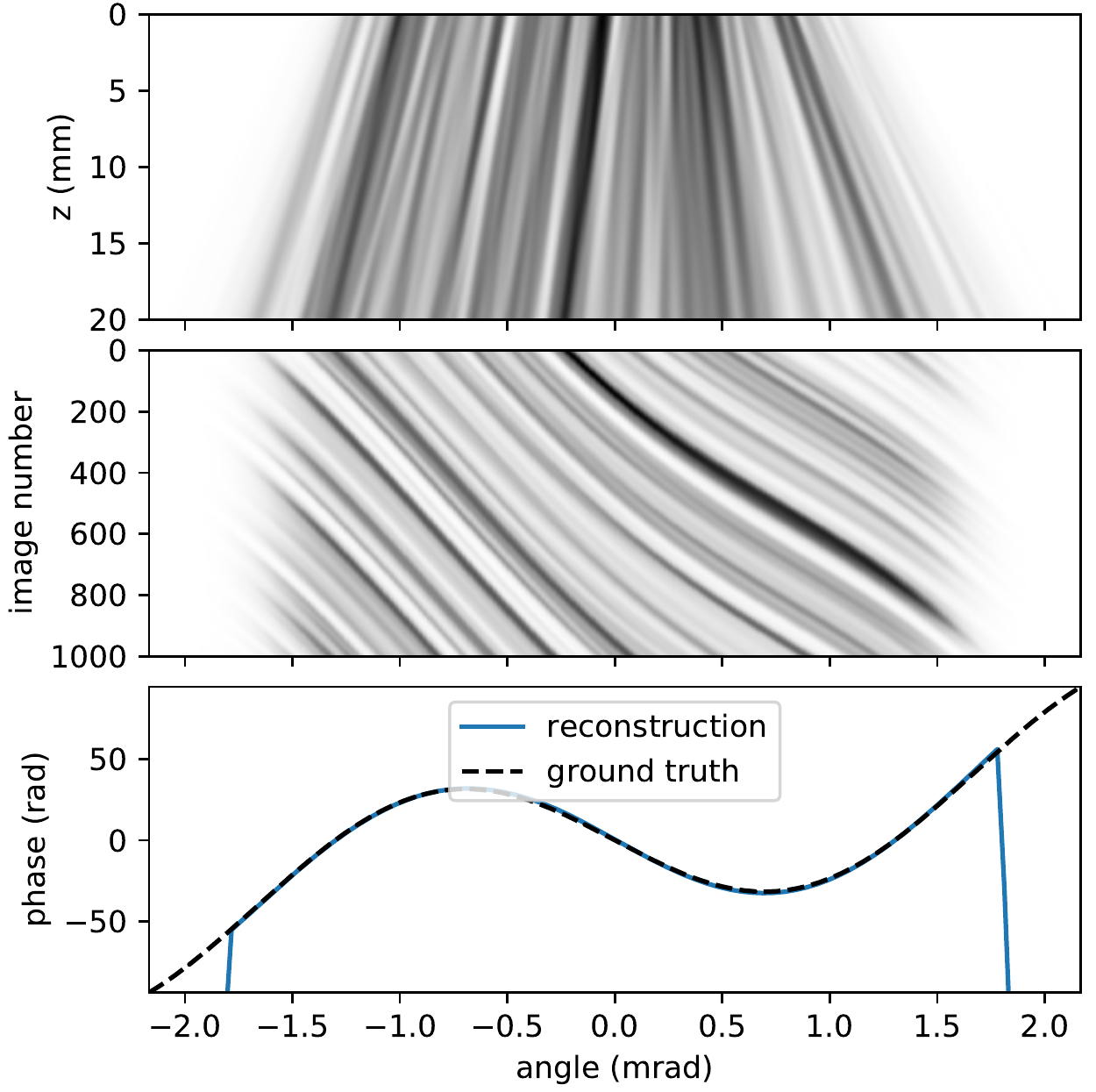}
\caption{\textbf{Top}: intensity of the wavefront propagating from the sample ($z=0$) to the detector plane ($z=20\;$mm).
The linear colour scale ranges from 0 (white) to the maximum value (black). 
\textbf{Middle}: stack of the 1D images recorded as the sample is scanned across the wavefield (to the right). The colour scale is the same as in top. 
\textbf{Bottom}: reconstructed and input phase aberrations in the detector plane. See text for further details.}
\label{fig:recon}
\end{figure}

The simulated sample is similar to that shown in Fig. \ref{fig:XST}. It was constructed 
in Fourier space with a Gaussian intensity profile and random phases at each pixel. 
The real-space object is thus complex valued, so that rays passing through the sample 
will be both absorbed and deflected in angle. The intensity of the illumination profile, 
in the plane of the detector, was formed by setting $W$ equal to a top 
hat function filtered with a Gaussian kernel. This filter produces a smooth tapered fall-off
in the intensity near the edges of the beam that helps to avoid aliasing artefacts during 
numerical propagation of the wavefront. The phase profile, $\Phi$, was constructed with 
the quadratic function $\pi x^2 / (\lambda (z_1+z))$, where $\lambda = 1.2\;$nm ($1\;$keV), 
$z=20\;$mm and $z_1=40\;$mm, so that the focal plane of the illumination is upstream of the sample in the top panel of Fig. \ref{fig:recon} by a distance that is twice that of the sample to detector distance. This leads to an average magnification factor of $1.5$. 
In addition to this, a sinusoidal phase profile was added to the phase in order to simulate the result of 
aberrations in the lens system, this can be seen as the dashed black line in the bottom panel of Fig. \ref{fig:recon}. 
The intensity of the wavefront, $I(x, z)$, propagating from the exit surface of the sample to the detector plane 
is shown in the top panel. Upon close inspection, one can see that the intensities in the plane of the 
detector are non-trivially related to those in the exit surface of the sample. As such $I(x, 0)$ cannot be
constructed from $I(x, z)$ by a scaling in $x$ (magnification) or indeed by any geometric mapping. We make the point again, 
that in PXST the ``reference'' image is not the sample transmission profile, rather, it is the intensity profile 
one would have observed on a detector placed a distance $\bar{z}=zz_1/(z_1+z)=13.3\;$mm downstream of the sample illuminated 
by a plane wave. It is the geometric mapping between the reference image (not the sample transmission) and the recorded images that is used to reconstruct the phase profile of the illumination. 

The advantage of this 1D example is that one can visualise the entire dataset in a single 2D image. 
In the middle panel of Fig. \ref{fig:recon} the 1D images formed on the detector, as the sample is 
scanned across the wavefield, are displayed as an image stack. Along the vertical axis is the image
number and the horizontal axis is in angle units, which are the angles made from the point source 
to each pixel in the image. It is seen that this image stack consists of a series of lines that appear to flow towards positive angles as the image number increases. These are the features in the image that can be obviously tracked through the stack.
In this representation, the gradient of the lines at each diffraction angle are proportional the 
local wavefront curvature. For example, at a diffraction angle of $\approx -0.75\;$mrad, the 
wavefront aberrations have a negative curvature and so features at this point in the wavefield
are demagnified with respect to the mean. At a diffraction angle of $\approx 0.75\;$mrad, the 
opposite is true (with a greater magnification) and the line gradients are shallow with respect 
to those at $\approx -0.75\;$mrad. In addition to variations in the geometric magnification, 
the wavefront aberrations also locally adjust the effective propagation distance of the speckles. 
This is a non-geometric effect and (unlike the local variations in the magnification) is
not accounted for by the PXST reconstruction algorithm. 
For the current example, we have deliberately set the aberrations such that the local magnification 
and effective propagation distance vary by a significant fraction across the wavefield. This allows
for their effect to be clearly observed in the simulated data, but also leads to some errors in the
phases. 

The reconstructed phase profile, after 30 iterations of the PXST algorithm, is shown as the blue
line in the bottom panel of Fig. \ref{fig:recon}. The pedestal, tilt and defocus terms have been
removed prior to display, to allow the sinusoidal aberration profile to be clearly visualised. 
Near the edges of the illumination, $W\approx 0$ and the phases could not be determined (as expected). 
Apart from this, the differences between the ground truth and reconstructed phase profile 
($0.1\;$rad root mean squared error) are too small to see in this plot but are still much greater 
the theoretical lower limit of $\approx 0.0001$ (this limit is defined in section \ref{sec:res}) -- due to 
strength of the aberrations (as described in the previous paragraph).

\section{Uniqueness}\label{sec:uniqueness}

\subsection{Pedestal and tilt terms are unconstrained in the illumination phase profile.}
\noindent In general, the solution to the target function in Eq. \ref{eq:calc_err} does 
not constrain terms proportional to $1$, $x$ and $y$ in the recovered phase 
profile. These terms are sometimes referred to as the ``pedestal'' and ``tilt'' 
components of the pupil function. To see this, consider a phase profile $\Phi' 
= c + \textbf{d}\cdot\textbf{x} + \Phi$, where $\Phi$ corresponds to the true phase profile in the 
plane of the detector and where $c$ and $\textbf{d}$ are constants. This leads to the 
phase gradients $\nabla \Phi' = \textbf{d} + \nabla \Phi$. Substitution into Eq. \ref
{eq:st} yields
\begin{align}
 I'_n(\textbf{x}) &= W(\textbf{x}) I_\text{ref}(\textbf{x} - \frac{\lambda z}{2\pi} \nabla \Phi'(\textbf{x}) 
                      - \Delta \textbf{x}_n, \bar{z}), \\
         &= W(\textbf{x}) I'_r(\textbf{x} - \frac{\lambda z}{2\pi} \nabla \Phi(\textbf{x}) 
                      - \Delta \textbf{x}_n , \bar{z}), \\
         &= I_n(\textbf{x}) \text{ for } I'_r(\textbf{x}, \bar{z}) 
         = I_\text{ref}(\textbf{x} - \frac{\lambda z}{2\pi}  \textbf{d}, \bar{z}),               
\end{align}
independent of the pedestal term. Also, the tilt in the phase profile has 
produced a shift in the reference image, which is generally undetectable unless 
the position of a feature in the object is known \textit{a priori} with respect to the detector. 

Alternatively, we could have absorbed the term $\frac{\lambda z}{2\pi} \textbf{d}$ as 
a constant offset in the sample translation vectors $\Delta \textbf{x}_n' = \Delta \textbf{x}_n + 
\frac{\lambda z}{2\pi} \textbf{d}$. We note that the tilt terms are typically 
unconstrained in speckle tracking techniques, since it is common to allow for 
an overall offset in the sample or detector positions.

\noindent \subsection{Speckle patterns of sufficient density are required for a unique 
solution to exist.}
\noindent Clearly, in the extreme case where no speckles are recorded then the phase is 
completely unconstrained, so that
\begin{align}
 I'_n(\textbf{x}) &= W(\textbf{x}) I_\text{ref}(\textbf{x} - \frac{\lambda z}{2\pi} \nabla \Phi'(\textbf{x}) 
                      - \Delta \textbf{x}_n, \bar{z}), \\
         &= I_n(\textbf{x}) \text{ for } I_\text{ref}(\textbf{x}, \bar{z}) = 1
\end{align}
and for any $\Phi'$. This condition could also be reached in the limit where 
the fringe visibility of the speckle pattern approaches zero. The requirement 
for adequate fringe visibility is a point which is emphasised by \citeasnoun{Zanette2014} as well as many others in the field \cite
{Zdora2018}. Of course, the above condition could also be reached for any sub-domain of $\textbf{x}$. If, for example, $I_n(\textbf{x}') = W(\textbf{x}')$ for all $n$, then the phase 
terms are unconstrained at the points $\textbf{x}'$. This suggests a more general 
(necessary) condition for a unique solution to exist: that a speckle of 
sufficient contrast must be observed at least once at each position in the 
image. As an example, the Hartmann sensor discussed in section \ref{sec:background} does not satisfy this condition and, as such, it is necessary to interpolate values of $\Phi$ between the mask holes, rendering the method insensitive to high order aberrations that lead to rapid variations in $\Delta \Phi$. We should note that this is not always an issue, especially in cases where the low order aberrations of the wavefront are of primary concern, such as when we wish to correct them by some means, or when the low order aberrations are dominant and dominate the imaging performance of the optic. 

In another extreme, the phase is also unconstrained for $N=1$, when only a 
single image has been recorded. This is because the unknown $I_\text{ref}$ can be 
adjusted to accommodate any $\Phi'$:
\begin{align}
 I'_0(\textbf{x}) &= W(\textbf{x}) I_\text{ref}(\textbf{x} - \frac{\lambda z}{2\pi} \nabla \Phi'(\textbf{x}) 
                      - \Delta \textbf{x}_0, \bar{z}), \\
         &= I_0(\textbf{x}) \\
 \text{ for } I_\text{ref}(\textbf{x}, \bar{z}) &= 
           [I'_0 / W](\textbf{x} + \frac{\lambda z}{2\pi} \nabla \phi'(\textbf{x}) + \Delta \textbf{x}_0),
\end{align}
and for any $\Phi'$. This situation has arisen because the observed speckles 
are modelled as a function of $\nabla \Phi$ and $I_\text{ref}$, both of which are 
refined in the PXST method. So if a given speckle is observed only once, at a 
location $\textbf{x}'$, and the phase gradient at $\textbf{x}'$ is not otherwise constrained, 
then multiple solutions for $\nabla \Phi$ and $I_\text{ref}$ exist. The above two 
considerations suggest the more general (necessary) conditions for a unique solution to 
exist:
\vskip 0.1cm
\noindent\textit{
A speckle of sufficient contrast must be observed at least once at each 
position in the image, and, this speckle must be observed at least twice and at 
different positions in the image.
}
\vskip 0.1cm
\noindent Note that the above constraint does not require that every speckle 
must be observed more than once. 

The easiest way to satisfy this condition is to use a sample that produces a 
dense, high contrast array of speckles on the detector, such as a diffuser. 
With such a sample, the above constraint may be satisfied with just two images 
(i.e. for $N\ge2$), provided that the sample step size is not greater than half 
the illuminated region of the sample along the direction of the step. 

However, it is possible for a unique solution to exist even when the sample 
produces only a single observable speckle in each image. In this case, the 
above condition can be satisfied by scanning the sample such that this speckle 
is observed at each point in the image. This is a far less efficient means for 
wavefront sensing than using a diffuser. But this generality allows for nearly 
any object, such as the Siemens star in Fig. \ref{fig:overview} or a Hartmann mask, to be used as a wavefront sensing device. 

\noindent \subsection{Ambiguities can arise due to unknown sample positions.} 
\noindent If the sample translation vectors ($\Delta \textbf{x}_n$) are unknown, then there exists 
a family of solutions to Eq. \ref{eq:st}, with each solution corresponding to a 
set of translation vectors related by an affine transformation. Consider a set 
of translation vectors $\Delta \textbf{x}'_n = \textbf{x}_0 + \textbf{A} \cdot \Delta \textbf{x}_n$, where $\textbf{x}_0$ is an overall offset and the dot 
product is between the $2\times 2$ linear transformation matrix $\textbf{A}$ and the true sample 
translation vectors. As described previously, any overall offset in the 
translation vectors generates a corresponding offset in the reference 
image and a tilt term in the recovered phases. So, neglecting the offset term, 
we generate this family of solutions by the substitution $\Delta \textbf{x}_n = \textbf{A}^{-1}
\cdot \Delta \textbf{x}_n'$ into Eq. \ref{eq:st}:
\begin{align*}
 I_n(\textbf{x}) &= W(\textbf{x}) I_\text{ref}(\textbf{x} - \frac{\lambda z}{2\pi} \nabla \Phi(\textbf{x}) 
                      - \textbf{A}^{-1} \cdot \Delta \textbf{x}_n', \bar{z}), \\
         &= W(\textbf{x}) I_\text{ref}\left(\textbf{A}^{-1} \cdot 
         \left[\textbf{A}\cdot \left(\textbf{x} - \frac{\lambda z}{2\pi} \nabla \Phi(\textbf{x})\right) 
                                    - \Delta \textbf{x}_n'\right], \bar{z}\right), \\
         &= W(\textbf{x}) I_\text{ref}'(\textbf{x} - \frac{\lambda z}{2\pi} \nabla \Phi'(\textbf{x}) 
                                    - \Delta \textbf{x}_n', \bar{z})
\end{align*}
where
\begin{align}
 \Delta \textbf{x}'_n      &= \textbf{A} \cdot \Delta \textbf{x}'_n, \nonumber \\
 I_\text{ref}'(\textbf{x}, \bar{z}) &= I_\text{ref}(\textbf{A}^{-1}\cdot \textbf{x}, \bar{z}), \text{ and } \nonumber\\
 \nabla \Phi'(\textbf{x})  &= \textbf{A}\cdot \nabla \Phi(\textbf{x}) 
                     + \frac{2\pi}{\lambda z}\left(\textbf{x} - \textbf{A}\cdot \textbf{x} \right). \label{eq:rotPhi} 
\end{align}
If, on the other hand, the true sample translation vectors are given by the 
input values of $\Delta \textbf{x}_n$, but with small random offsets of mean 0, then the 
true solution for the phase and reference image can be recovered from the 
retrieved values by removing the effect of any affine transformation that may 
have arisen during the reconstruction. This can be accomplished by minimising $
\sum_n |\Delta \textbf{x}^\text{out}_n - A\cdot \Delta \textbf{x}^\text{in}_n|^2$ with respect to $A$, where $
\Delta \textbf{x}^\text{in}_n$ and $\Delta \textbf{x}^\text{out}_n$ are the input and output 
values of $\Delta \textbf{x}_n$ respectively, then generating the corresponding 
solutions for $\nabla \Phi'$ and $I'_r$ from the above equations. This 
situation can arise, for example, due to small relative errors in the 
translation of a stepper motor or from the pointing jitter of an XFEL pulse.

\noindent \subsection{An unknown rotation of the sample stage axes with respect to the 
detector axes can be corrected.}
\noindent A common systematic error for the input sample positions, is an overall 
rotation of the axes of the sample translation stages with respect to the pixel 
axes of the detector. In this case the linear transformation matrix reduces to 
the rotation matrix
\begin{align}
 \textbf{A} \rightarrow \textbf{R}(\theta) = 
 \begin{bmatrix}
 \cos{\theta} & -\sin{\theta} \\
 \sin{\theta} & \cos{\theta}
 \end{bmatrix}.
\end{align}
Here, we can make use of the fact that in general $\nabla \Phi'$ of Eq. \ref{eq:rotPhi}
is \textit{not} irrotational for $\theta\neq 0$ and $\textbf{A} = \textbf{R}(\theta)$. If $(u, v) \equiv \nabla 
\Phi^\text{out}$ then we can demand that the vector field $\textbf{R}^{-1} \cdot (u, v)$ be irrotational. With $\textbf{R}^{-1} = \textbf{R}(-\theta)$ and Eq. \ref
{eq:irrotational}, we have
\begin{multline}
 %
 \theta = \text{argmin}_\theta\bigg[\bigg( 
          \frac{\partial}{\partial y}(u \cos{\theta} + v\sin{\theta}) \\
          - \frac{\partial}{\partial x}(-u \sin{\theta}  + v\cos{\theta}) \bigg)^2 \bigg]
\end{multline}
where the derivatives with respect to $x$ and $y$ are evaluated numerically. 
However, if the irrotational constraint was enforced during the reconstruction, 
then the recovered $\Delta \textbf{x}^\text{out}_n$ is free of the erroneous 
rotation and no further analysis is required. 

\noindent \subsection{The raster grid pathology produces artefacts for lattice-like 
sample translations.} 
\noindent A common annoyance encountered in ptychography is the so called ``raster grid 
pathology'' \cite{Thibault2009}. The raster grid pathology arises when reconstructing both 
the illumination and sample profiles from diffraction data acquired while the 
sample is scanned along a regular grid. In that case, the recovered 
illumination and sample transmission functions may be modulated by any 
function, so long as it is periodic on a lattice of points upon which all of 
the sample positions lie. 

In many cases, the governing equation for a ptychographic reconstruction is 
given by
\begin{align}\label{eq:pty}
 I_n(\textbf{q}) &= \bigg|\mathcal{F}\left[ T(\textbf{x}-\Delta \textbf{x}_n) p(\textbf{x}, 0) \right]\bigg|^2
\end{align}
where $\mathcal{F}\left[\cdot\right]$ is the Fourier transformation operator 
over the transverse plane and represents the propagation of the exit-surface 
wavefront $\psi_n(\textbf{x}) \equiv T(\textbf{x}-\Delta \textbf{x}_n) p(\textbf{x}, 0)$ to the detector (in the 
far-field of the sample). If the sample is translated along a regular grid, for 
example, with step size $\textbf{d}$ then $\Delta \textbf{x}_n = \textbf{n} \cdot \textbf{d}$, where the vector $\textbf{n} = (i_n, j_n)$ is the 2D lattice index corresponding to the $n$\textsuperscript{th} image (for integer $i_n$ and $j_n$). 
If we make the substitution $p'(\textbf{x}, 0) = f(\textbf{x}) p(\textbf{x}, 0)$ into the above equation, 
then we have 
%
\begin{align*}
 \psi_n(\textbf{x}) &= T'(\textbf{x} - \textbf{n}\cdot \textbf{d}) f(\textbf{x}) p(\textbf{x}, 0) \\
 \therefore T'(\textbf{x}) &= \frac{\psi'_n(\textbf{x}+\textbf{n}\cdot\textbf{d})}{p(\textbf{x}+\textbf{n}\cdot\textbf{d}, 0) f(\textbf{x}+\textbf{n}\cdot\textbf{d})} 
 = \frac{T(\textbf{x})}{f(\textbf{x})} \\
 \text{if } f(\textbf{x} \pm \textbf{n}\cdot\textbf{d}) &= f(\textbf{x}) \text{ for all } \textbf{n}
\end{align*}
The raster grid pathology can be avoided by ensuring that the sample scan 
positions lack any translational symmetry, i.e. by scanning the sample in non-regular patterns, for example in a spiral grid, or by adding a random 
offset to every grid position \cite{Fannjiang2018}. 
Given that Eq. \ref{eq:pty} is nothing but a special case of the Fresnel 
integral in Eq. \ref{eq:Fr} (from which the speckle tracking approximation is 
derived) it is natural to consider whether or not the same pathology applies 
here. 

One can show that a similar pathology does indeed arise in the present case. 
Here, the illumination's intensity is constrained during the reconstruction, so 
instead we make the substitution $p'(\textbf{x}, z) = p(\textbf{x}, z) e^{ig(\textbf{x})}$, which is equivalent to $\Phi'(\textbf{x}) = 
\Phi(\textbf{x}) + g(\textbf{x})$, into Eq. \ref{eq:st}: 
%
\begin{align*}
 I_n(\textbf{x}) &= W(\textbf{x}) 
           I'_r(\textbf{x} - \frac{\lambda z}{2\pi} \nabla \Phi'(\textbf{x}) - \textbf{n}\cdot\textbf{d}, \bar{z}), \\ 
        &= W(\textbf{x}) I'_r(\textbf{x} - \frac{\lambda z}{2\pi} 
                           (\nabla \Phi(\textbf{x}) + \nabla g(\textbf{x}))  - \textbf{n}\cdot\textbf{d}, \bar{z}), \\ 
        &= W(\textbf{x}) I'_r((\textbf{x} - \nabla g(\textbf{x} - \textbf{n}\cdot\textbf{d})) - 
                     \frac{\lambda z}{2\pi} \nabla \Phi(\textbf{x}) - \textbf{n}\cdot\textbf{d}, \bar{z}), \\ 
 \therefore I'_r(\textbf{x}) &= I_\text{ref}(\textbf{x} + \frac{\lambda z}{2\pi} \nabla g(\textbf{x})),\\
 \text{if } \nabla g(\textbf{x}) &= \nabla g(\textbf{x} \pm \textbf{n}\cdot\textbf{d}) \text{ for all } \textbf{n}.
\end{align*}
So, rather than modulating the reference image with a periodic function, the 
pathology here creates a geometric distortion of the reference image.



\section{Angular Sensitivity and Imaging Resolution}\label{sec:res}


\noindent \subsection{The resolution of the reference image is given by the demagnified effective pixel size.}
\noindent As discussed in section \ref{sec:method}, the Fresnel scaling theorem states that the projection image of thin sample formed by a point like source of coherent light produces a magnified and defocused image of the sample. Similarly, in PXST, the ``reference image'' is an idealised image that would have been formed if the illumination were plane-wave (i.e. with a flat phase profile), the detector were placed a distance $\bar{z}$ from the plane of the sample, the detector extended over the entire illuminated region of the sample, and the physical pixel size ($\sigma_\text{det}$) was reduced by the magnification factor $M$. Assuming that the speckle tracking approximation holds, and that the aggregate signal-to-noise level is high, then the resolution of such an image is given by the de-magnified effective pixel size of the detector ($\sigma_\text{ref}$). 

The effective pixel size ($\sigma_\text{eff}$) can be much smaller than $\sigma_\text{det}$ due to sub-pixel interpolation, which is employed when registering a speckle across many images. We have found, as others have noted \cite{Berujon2012}, that sub-pixel interpolation can lead a reduction in the effective pixel size by a factor of $10$ or more depending on the point-spread function of the detector, the contrast of the speckles, the signal-to-noise per image and the total number of images. On the other hand, effects such as the finite source size of the x-rays will tend to blur-out speckles and increase the effective pixel size. 

Consider the imaging geometry depicted in Fig. \ref{fig:stem}: for an incoherent source of x-rays with a Gaussian angular distribution given by $\exp{[-\theta^2/2\sigma_s^2]}$, where $\theta$ is the angle made by a ray pointing from the incoherent source point to the lens aperture. Then the image recorded in the detector plane will be given by
\begin{align*}
 I(\textbf{x}, z; \sigma_s) &\approx \left( \frac{z_1}{zf2\pi \sigma_s^2} e^{-\frac{\textbf{x}^2}{2}\left(\frac{z_1}{z f \sigma_s}\right)^2 } \right) \otimes I(\textbf{x}, z),
\end{align*}
where $\otimes$ is the convolution operator, $f$ is the focal length of the lens and $I(\textbf{x}, z)$ is the intensity of the wavefront in the plane of the detector for an on-axis point source of light. Therefore, the observed speckles will be broadened by a factor $\sigma_s zf / z_1$. In this r\'{e}gime, the effective pixel size is given by
\begin{align}
 \sigma_\text{eff} &= \delta_\text{pix} \times \left( \sigma_\text{det} + \sigma_s \frac{z f}{z_1} \right), \\
 \sigma_\text{ref} &= \frac{\sigma_\text{eff}}{M}, \label{eq:sref}
\end{align}
where we have used the symbol $\delta_\text{pix}$ to represent the fractional reduction in the effective pixel size due to numerical interpolation. For example, with a physical pixel size of $\sigma_\text{eff} = 50\;\mu$m, a fully coherent wavefield and $\delta_\text{pix}=1$, the de-magnified pixel size for the example shown in Fig. \ref{fig:stem_im} (left and middle columns) would be $\sigma_\text{ref}=25\;$nm and for the right column (with the sample $10\;\mu$m from the focus) $\sigma_\text{ref}=5\;$\AA. 


\noindent \subsection{The sample position that maximises the angular sensitivity of the wavefront, in the plane of the sample, depends on the source coherence width.}

\noindent The smallest resolvable angular deviation of the wavefront, in the plane of the sample, is given by the arctangent of the smallest resolvable displacement of a speckle over the distance between the sample to the detector pixel array
\begin{align}
 \Delta \Theta_\phi &= \arctan{\frac{\sigma_\text{eff}}{z}} \approx \frac{\sigma_\text{eff}}{z}, \label{eq:dtphi}
\end{align}
where the small angle approximation (employed here) almost certainly holds for any applications of interest. The smallest resolvable increment in the phase gradient is thus $\Delta (\nabla \phi) = \frac{2\pi}{\lambda} \Delta \Theta_\phi$. 


For the imaging geometry depicted in Fig. \ref{fig:stem}, the optimal position for the detector will be given by the furthest distance from the focus such that the footprint of the illumination is contained within the pixel array, as this maximises the sampling frequency of the wavefield. This raises the question: how far should one place the sample from the focus in order to maximise the angular resolution (minimise $\Delta \Theta$)? To answer this, let us keep the focus-to-detector distance ($z_t$) fixed, so that $z_t = z_1 + z$ and minimise $\Delta \Theta_\phi$ with respect to $z_1$. Inserting $M= z_t / z_1$ and Eq. \ref{eq:sref} into Eq. \ref{eq:dtphi} and minimising yields
\begin{align*}
 \Delta \Theta_\phi &= \frac{\sigma_\text{det}}{z_t} (1+\sqrt{\frac{f\sigma_s}{\sigma_\text{det}}})^2, \\
  \text{ for} \;\; z_1 &= \frac{z_t}{1 + \sqrt{\frac{\sigma_\text{det}}{f\sigma_s}}}.
\end{align*}
As the focus-to-sample distance is reduced, the magnification factor increases (improving the angular sensitivity), while at the same time the deleterious effects of the finite source size increase (deteriorating the angular sensitivity). The above value for $z_1$ represents the optimal compromise between these two effects. 

\noindent \subsection{The sample position that maximises the angular sensitivity of the wavefront, in the plane of the detector, is in the focal plane where the magnification is greatest.}
\noindent The angular resolution for the wavefield in the plane of the detector ($\Delta \Theta_\Phi$) is not, in general, the same as the angular resolution in the plane of the sample. This is because of the difference in extent between the effective pixel size and the demagnified effective pixel size that arises due to divergent illuminating wavefields. $\Delta \Theta_\Phi$ is given by
\begin{align}
 \Delta \Theta_\Phi &= \frac{\sigma_\text{ref}}{z}. \label{eq:dtPhi}
\end{align}
For a fixed focus-to-detector distance, once again $z = z_t-z_1$, and we have
\begin{align}
 \Delta \Theta_\Phi &= \delta_\text{pix} \times \left( \frac{z_1 }{z_t} \sigma_\text{det} + \sigma_s \right). \label{eq:dtPhi2}
\end{align}
An interesting feature of the above equation is that the effect of the source incoherence $\delta_\text{pix}\sigma_s$ does not vary with $z_1$; as $z_1$ is decreased the increase in the magnification factor leads to a corresponding decrease in the effective pixel size due to the finite source size, but this
is exactly balanced by the reduction in the angle subtended by the demagnified effective pixel size. Therefore, the optimal position for the sample, in order to minimise $\Delta \Theta_\Phi$, is as close to the focal plane as possible. However, this can be a dangerous limit to approach, as the speckle tracking approximation will begin to break down due to the rapidly varying illuminating wavefield -- at this point, it would be necessary to employ a fully coherent model for the wavefront propagation, for example, by switching to far-field ptychography. Additionally, it can be beneficial to increase the focus-to-sample distance for practical reasons; for example, increasing $z_1$ provides a larger field-of-view of the sample in each image which aids in positioning the region of interest of the sample with respect to the illuminating beam and, typically, increases the speckle visibility. For these reasons, it can be beneficial to approach the limit where
\begin{align}
  \frac{\sigma_\text{det}}{M} &\approx \sigma_s
  \quad \text{ or }\quad z_1 \approx z_t \frac{\sigma_s}{\sigma_\text{det}}.
\end{align}
Here $z_1$ has been increased until the demagnified pixel size is approximately equal to the demagnified feature size one would observe from a point-like object due to incoherence alone. This represents the transition between modes where $\Delta \Theta_\Phi$ is dominated by the detector pixel size (larger $z_1$) and by the finite source size (smaller $z_1$).

\noindent \subsection{Although the angular sensitivity of the wavefields in the sample and detector planes differ, the phase sensitivities are equal and are both minimised by maximising the magnification.}
\noindent The difference between $\Delta \Theta_\phi$ and $\Delta \Theta_\Phi$ may at first appear to be a curious asymmetry. However, the phase sensitivity in the plane of the sample and the detector are in fact equal for a given sample position. For divergent illumination, the angular distribution of the wavefield in the sample plane is larger than that of the wavefield in the detector plane by a factor of $M$. On the other hand, the sampling frequency of the wavefield in the sample plane is also larger by the factor $M$. So, when propagating uncertainties in the angular distributions to the integrated phase profiles, these two effects cancel. 

Recall the relationship between the integrated phase profile and the angular distribution of the wavefield in Eq. \ref{eq:int_angle}. Uncertainties in the angular distribution are thus scaled by the step size in $x$ after integration, so that
\begin{align}
 \Delta \phi &\approx \sigma_\text{ref} \times \frac{2\pi}{\lambda} \Delta \Theta_\phi
 =   \sigma_\text{ref} \times \frac{2\pi}{\lambda}\frac{\sigma_\text{eff}}{z}, \\
 \Delta \Phi &\approx \sigma_\text{eff} \times \frac{2\pi}{\lambda} \Delta \Theta_\Phi
 =  \sigma_\text{eff} \times \frac{2\pi}{\lambda}  \frac{\sigma_\text{ref}}{z}. \label{eq:dPhi}
\end{align}
Therefore, with $\sigma_\text{ref} = \sigma_\text{eff} / M$, we have that $\Delta \phi = \Delta \Phi \approx \frac{2\pi}{\lambda} \sigma_\text{eff}^2/M$ and both are minimised by placing the sample as close as possible to the focal plane, as described in the previous subsection.

\section{Discussion and conclusion}\label{sec:discussion}


We have presented a modified form of the speckle tracking approximation, valid to second-order in a local expansion of the phase term in the Fresnel integral. This result extends the validity of the speckle tracking approximation, thus allowing for greater variation of the unknown phase profile and thus for greater magnification factors when the wavefield has a high degree of divergence (such as that produced by a high numerical aperture lens system) or, when imaging a sample in the differential configuration of XST, to allow for greater phase variation across the transmission function of the sample (such as that produced by a thick specimen). We suggest that this approximation can be used, with little modification, in many of the existing XST applications and suggest such a modification for the UMPA approach.

We have also presented the PXST method, a wavefront metrology tool capable of dealing with highly divergent wavefields (like XSS) but unlike XSS, the resolution does not depend on the step size of the sample translations transverse to the beam. Coupled with a high numerical aperture lens, PXST provides access to nanoradian angular sensitivities as well as highly magnified views of the sample projection image. With a suitable scattering object, which in this case is the sample itself, a minimum of two images are required although more images will improve robustness and resolution. 

We must emphasise that it is only the projection image of the sample that is recovered. The phase and transmission profile of the sample must be inferred from the projection image via standard techniques \cite{Wilkins2014}. This is in contrast to other methods that provide multiple modes of imaging of the sample, such as the transmission, phase and the so called ``dark-field'' profiles. What distinguishes PXST from these methods, is that the sample image is obtained in addition to the wavefield phase in the absolute configuration of XST; that is, both can be obtained from a single scan series of the sample.  

A further application of this method is to use it as an efficient prior step to Fourier ptychography, by recording images out of focus. The recovered illumination and sample profiles can be used as initial estimates for a Fourier ptychographic reconstruction. Experimentally, this additional step can be achieved simply by moving the sample towards the focal plane of the lens. In some cases, this additional step would not even be required, so that speckle tracking followed by ptychography could be performed on the same dataset.

For experimental results utilising the PXST method, see \cite{Morgan2019a}. These results are based on a campaign of measurements for the development of high numerical aperture wedged multi-layer Laue lens systems.

\section{Acknowledgements}

We would like to acknowledge Timur E. Gureyev, for proof reading the manuscript and for fruitful discussion on the theory of x-ray wave propagation. We also acknowledge Chufeng Li for additional proof reading. Funding for this project was provided by: the Australian Research Council Centre of Excellence in Advanced Molecular Imaging (AMI) and the Gottfried Wilhelm Leibniz Program of the DFG.

\bibliographystyle{iucr}
\bibliography{speckle_theory}

\newpage

\appendix 

\section{Derivation of the Speckle Tracking Approximation}\label{sec:approx}

The Fresnel integral of Eq. \ref{eq:Fr} is often referred to as a point 
projection mapping. This is because, when the Fresnel number is $\gg 1$, the dominant contributions to the integral 
typically arise from values of the sample transmission, $T(x')$, around the point $x$ (i.e. for 
$x'\approx x$). At this point, the phase term $\pi (x-x')^2/\lambda z$ has a 
spatial frequency $q_\text{F} = (x-x')/\lambda z \approx 0$. For $x'$ far from 
$x$, the phase term causes the integrand to oscillate rapidly between $\pm 
T(x')$. If $T(x')$ is bandwidth limited, with a maximum spatial frequency of 
(say) $q_\text{max}$, then, for a sufficiently large $|x-x'|$ $q_\text{F} \ge q_\text{max}$ and 
successive oscillations of the integrand, caused by the phase term, will occur 
at roughly the same values of $T(x')$ and will thus cancel each other in the 
integration. 

However, in the Fresnel integral of Eq. \ref{eq:I}, the modulation of $T(x')$ 
by $p(x', 0)=\sqrt{w}(x') e^{i\phi(x')}$ has generated an additional phase 
term, and we would now expect the dominant contribution to $I(x,z)$ from 
$T(x')$ to arise at values of $x'$ for which the integrand is smooth. To simplify the analysis, let us 
gather the phase terms of from the Fresnel exponent and the incident illumination into a global phase factor:
\begin{align}\label{eq:gdef}
  g(x, x') &= \phi(x') + \frac{\pi }{\lambda z}(x-x')^2,
\end{align}
so that the complex amplitude of the Fresnel integral becomes\footnote{For simplicity, 
the following analysis will be presented in 1D. The difference between one and two dimensions is mostly in the normalisation constants. At the end of this section we will generalise the result to 2D}
\begin{align}\label{eq:Fcoh}
  \psi(x,z) &= \frac{e^{2\pi i z/\lambda}}{\sqrt{-i\lambda z}} 
               \int T(x') p(x, x') e^{\frac{i\pi}{\lambda z} (x-x')^2 } dx', \\
  &= \frac{e^{2\pi i z/\lambda}}{\sqrt{-i\lambda z}} 
               \int T(x') \sqrt{w}(x') e^{i g(x, x')} dx'. \label{eq:Fcoh2}
\end{align}
Note that the global phase term, $g(x, x')$, does not contain any contribution from the phase of the transmission function $T(x')$. Without any prior knowledge of this phase term, our smoothness condition becomes
\begin{align}
 \frac{\partial g(x,x')}{\partial x'} &= \nabla \phi(x') - 
 \frac{2\pi}{\lambda z}(x-x') = 0. 
 %
\end{align}
For now, we will call the solution to this equation for $x$ given $x'$, $u^{-1}$, so that:
\begin{align}
 x &= u^{-1}(x') = x' + \frac{\lambda z}{2\pi} \nabla \phi(x'), \label{eq:g}
\end{align}
where $u^{-1}$ is the functional inverse of $u$, which is yet to be defined.
So, Eq. \ref{eq:I} now represents the point projection mapping 
$x' \rightarrow u^{-1}(x')$, rather than $x'\rightarrow x$. 
\begin{figure}
\includegraphics[width=8.88cm]{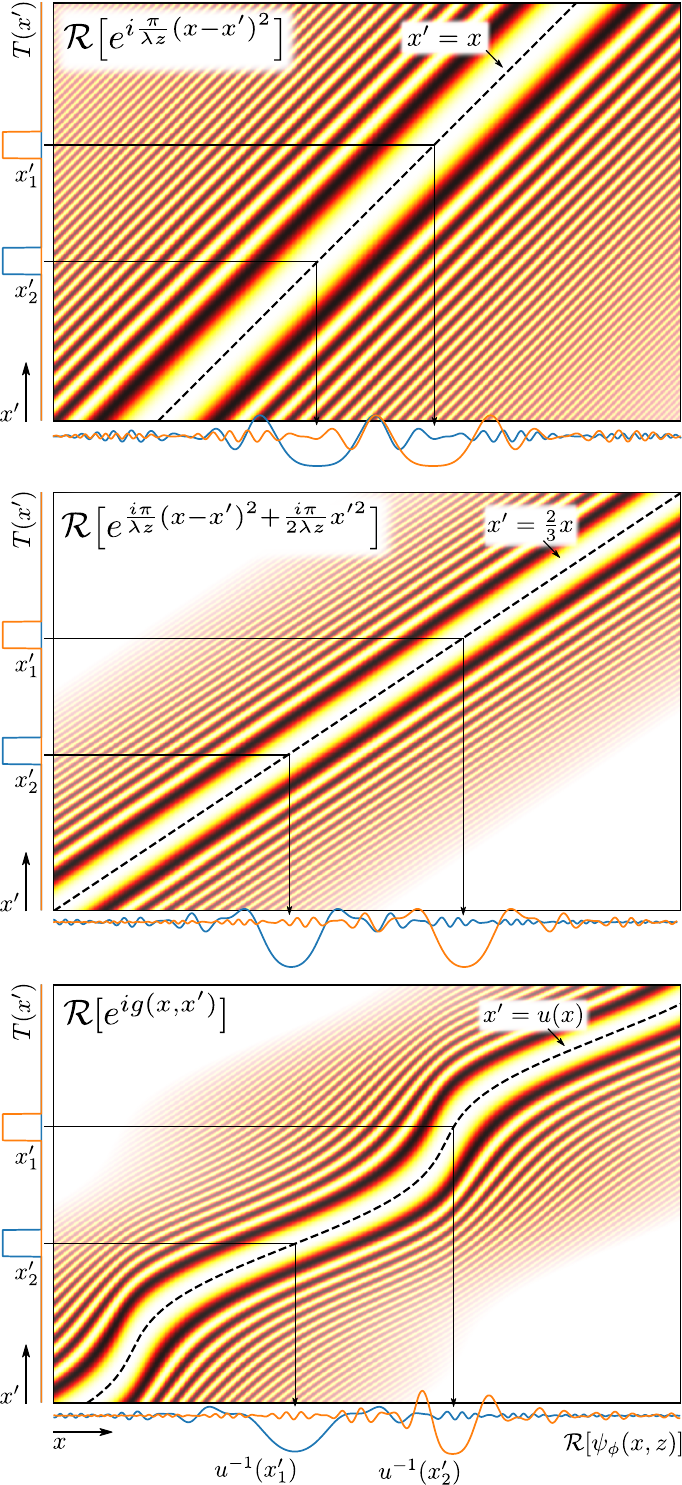}
\vspace*{-1.0cm}
\caption{Illustration of the Fresnel integral, with a modulating phase term, for two top hat functions. The colour maps display the real part of the exponential term, $\mathcal{R}\left[e^{g(x, x')}\right]$, with the same colour scale as in Fig. \ref{fig:Hartmann}. The alpha channel of the colour scale has been increased in regions where the spatial frequency of the exponent approaches the pixel size, so as to make transparent pixels that would otherwise be aliased. Phase terms that are constant with respect to $x$ have been subtracted before display, these terms would not effect the final intensity of the image and removing them more clearly shows the line $x'=u(x)$ where $\frac{\partial g}{\partial x'}=0$. See text for further details. } 
\label{fig:mapping}
\end{figure}
This point is 
illustrated in Fig. \ref{fig:mapping} with $\sqrt{w}(x')=1$ for three different values of $\phi(x')$. 
On the vertical axis of each panel, we plot two real top hat functions 
representing possible values for $T(x')$. In the two dimensional domain of the 
integrand, $T(x')$ is constant along the $x$-axis. The Fresnel integral is 
performed by extruding $T(x')$ along the horizontal axis, multiplying by 
$e^{i g(x, x')}$ then integrating along the vertical axis. The real part  
of this integral is illustrated along the horizontal axis of each panel. We 
can see that the centroids of the features, before and after the Fresnel 
integral, follow the point projection mapping defined by the dashed line 
$x'=u(x)$ which is defined by the condition $\frac{\partial g}{\partial x'}=0$. 
In the top panel $\phi(x')=0$, leading to $x' = u(x) = x$. This corresponds to Fresnel propagation with plane wave illumination. Therefore, the separation between the top hat functions in the sample plane are equal to the separation between the ``speckles'' produced by each top hat function. In the second panel $\phi(x') = \pi x'^2 / (2\lambda z)$, corresponding to divergent illumination that would arise from a point source of illumination, or an ideal lens system with an infinite numerical aperture. Here $x' = u(x) = 2x/3$ and, consistent with the Fresnel scaling theorem, this leads to both a geometric magnification (the speckles are separated by a greater distance than the top hats) and a change in the effective propagation distance (which can be observed in the more rapid oscillation of the exponential term). In the final panel, a sinusoidal phase term has been added to the phase from the middle panel. Here $u(x)$ does not have a simple form and both the effective propagation distance and the magnification vary with position along the the $x$-axis. 

This suggests the following modification to the approach outlined by 
Zanette \textit{et al.}: instead of expanding $\phi(x')$ and $\sqrt{w}(x')$ 
about $x$ in Eq. \ref{eq:I}, we should shift this expansion about the point 
$x'=u(x)$. In this way, the Taylor series expansion will be most accurate over
the domain of the integrand that contributes most to the integral. The $N$\textsuperscript{th}-order 
Taylor series expansion of $g(x, x')$ and $\sqrt{w}(x')$ about $x'=u(x)$ are 
given by
\begin{align} 
 g_N(x, x') &= \sum_{n=0}^{n=N} \frac{(x'-u(x))^n}{n!} 
             \frac{\partial^n g}{\partial x'^n} \rvert_{x, x'=u(x)}, \label{eq:gN} \\
 \sqrt{w}_N(x') &= \sum_{n=0}^{n=N} \frac{(x'-u(x))^n}{n!} 
             \frac{\partial^n w}{\partial x'^n} \rvert_{x'=u(x)}.\label{eq:wN} 
\end{align} 
Evaluating Eq. \ref{eq:gN} for $N=1$ and Eq. \ref{eq:wN} for $N=0$, we have
\begin{align*}
g_1(x, x') &= g(x, u(x)) + \frac{\partial g}{\partial x'} \rvert_{x, x'=u(x)} (x-u(x)) =  g(x, u(x)), \\
\sqrt{w}_0(x') &= \sqrt{w}(u(x)),
\end{align*}
where the $n=1$ term in the expansion of $g$ is zero by construction. With $g\approx g_1$ and $\sqrt{w}\approx\sqrt{w}_0$, Eq. \ref{eq:Fcoh} becomes
\begin{align*}
\psi(x, z) &\approx \frac{e^{2\pi i z/\lambda}}{\sqrt{-i\lambda z}} 
                    \sqrt{w}(u(x)) e^{i g(x, u(x))} \int T(x')  dx'.
\end{align*}
Unfortunately, the above expression completely fails to capture the 
physics upon which XST methods are based, i.e. the geometric mapping between $I$ and $I_\text{ref}$ defined by $\phi$. In our attempt to improve the accuracy of the speckle tracking approximation, the first-order expansion of $g$ about $x'=u(x)$ no longer depends on $x'$. Thus the integral over $x'$ in Eq. \ref{eq:Fcoh} has reduced to the term $\int T(x') dx'$.
\vskip 0.2cm
\noindent With the above result in mind, let us try the following approach: \\
\noindent\textit{
Instead of expanding $\phi(x')$ to first-order and $\sqrt{w}(x')$ to 
zeroth-order about the point $x'=x$, expand $g(x, x')$ to second-order and 
$\sqrt{w}(x')$ to zeroth-order about the point $x'=u(x)$.} \\
\noindent This approach leads to
\begin{align} 
 g_2(x, x') &= g(x, u(x)) +  \frac{1}{2} \bigg[ \nabla^2\phi(u(x)) + 
             \frac{2\pi}{\lambda z}\bigg] (x'-u(x))^2,
 \label{eq:ge}
\end{align} 
where, once again, the $n=1$ term for $g(x, x')$ is zero by construction in Eq. \ref{eq:g}. Substituting Eqs \ref{eq:ge} and $\sqrt{w}(x')\approx \sqrt{w}(u(x))$ into \ref{eq:Fcoh2}, then completing the square in the exponent we can recast the Fresnel integral in the following form
\begin{align} \label{eq:I2}
\psi(x, z) &\approx  
        \frac{e^{2\pi i z/\lambda}}{\sqrt{-i\lambda z}} w(u(x)) e^{i g(x, u(x))} 
        \int T(x') e^{\frac{i\pi}{\lambda z(u(x))} (x' - u(x))^2} dx' ,
\end{align}
where we have defined $z(x)$ as
\begin{align} 
    z(x) &\equiv \left[\frac{1}{z} + \frac{\lambda}{2\pi}\nabla^2 
                 \phi(x)\right]^{-1}.
    \label{eq:zx}
\end{align} 
One can interpret $z(x)$ as the propagation distance required to locally reproduce the diffraction features in $\psi(x, z)$ had the illumination been plane wave (i.e. $\phi(x') = 1)$. 

We remind the reader that the ``ref'' subscript refers to the wavefront that would have been formed with plane wave illumination, with $p(x, 0) = 1$. Here we define, $\psi_\text{ref}$ as the complex amplitudes corresponding to the Fresnel integral in Eq. \ref{eq:Fr}:
\begin{align}\label{eq:Fref}
  \psi_\text{ref}(x,z) &= \frac{e^{2\pi i z/\lambda}}{\sqrt{-i\lambda z}} 
               \int T(x') e^{\frac{i\pi}{\lambda z} (x-x')^2 } dx'.
\end{align}
Now $\psi(x, z)$ can be related to $\psi_\text{ref}(x, z)$ (where $I_\text{ref}=|\psi_\text{ref}|^2$) by the substitutions
\begin{align*} 
    x &\rightarrow u(x) & &\text{and} & z &\rightarrow z(u(x)).
\end{align*} 
Yielding 
\begin{align} 
\psi(x, z) &\approx e^{2\pi i (z-z(u(x)))/\lambda}  \times \label{eq:Fr2coh}\\
        & \qquad \sqrt{\frac{z(u(x))}{z}}  w(u(x)) e^{i g(x,u(x))}  \psi_\text{ref}(u(x), z(u(x))),  \nonumber \\
I(x, z) &\approx \frac{z(u(x))}{z} w(u(x)) I_\text{ref}(u(x), z(u(x))) 
\label{eq:Fr2}.
\end{align}
So far we have avoided a more explicit definition of the geometric mapping 
factor $u(x)$, it is currently defined by its inverse in Eq. \ref{eq:g}. A 
more meaningful definition can be obtained by the following consideration. 
Setting $T(x')=1$, Eq. \ref{eq:I2} represents 
the propagation of the incident beam through free space in the absence of the sample, so that
\begin{align*}
 p(x, z) &\approx \frac{e^{2\pi i z/\lambda}}{\sqrt{-i\lambda z}} w(u(x)) e^{i g(x, u(x))} 
        \int e^{\frac{i\pi}{\lambda z(u(x))} (x' - u(x))^2} dx' \\
         &= e^{2\pi i z/\lambda} \sqrt{\frac{z(u(x))}{z}} \sqrt{w}(u(x)) e^{i g(x, u(x))} \\
         &= e^{2\pi i z/\lambda} \sqrt{W}(x) e^{i\Phi(x)},
\end{align*}
where we have defined 
\begin{align}\label{eq:pyx} 
    W(x) &\equiv \frac{z(u(x))}{z} w(u(x)) & &\text{and} & \Phi(x) &\equiv g(x, u(x)),
\end{align} 
and $W(x)$ and $\Phi(x)$ are, respectively, the intensity and phase profiles 
of the undisturbed beam in the plane of the detector.  

The benefit of this calculation is that it provides an interpretation of the 
mapping function $u(x)$ in terms of the phase gradient of the illumination in 
the $z$ plane. To see this, we must perform a little more mathematical 
gymnastics. First, we explicitly evaluate $\Phi(x)$ in terms of the incident 
phase profile $\phi(x)$. Substituting $x'=u(x)$ into Eq. \ref{eq:gdef}, we 
have
\begin{align} 
    \Phi(x) &= \phi(u(x)) + \frac{\pi}{\lambda z} (x-u(x))^2.
    \label{eq:Phi}
\end{align}
Using the definition for $u^{-1}(x)$ in Eq. \ref{eq:g}, we can then evaluate
\begin{align} 
 \Phi(u^{-1}(x)) &= \phi(x) + \frac{\pi}{\lambda z} 
                    (x - x - \frac{\lambda z}{2\pi} \nabla \phi(x))^2 \nonumber \\
                 &= \phi(x) + \frac{\lambda z}{4\pi} (\nabla \phi(x))^2. 
 \label{eq:Phi2}
\end{align}
Taking the derivative of both sides of Eq. \ref{eq:Phi2} with respect to $x$ 
yields
\begin{align*} 
 \frac{\partial u^{-1}(x)}{\partial x} \nabla \Phi(u^{-1}(x)) 
 &=  \frac{\partial }{\partial x} 
 \left[\phi(x) + \frac{\lambda z}{4\pi} (\nabla \phi(x))^2\right], \\
 \left[1 + \frac{\lambda z}{2\pi} \nabla^2\phi(x)\right] \nabla \Phi(u^{-1}(x)) 
 &= \left[1 + \frac{\lambda z}{2\pi} \nabla^2 \phi(x)\right]\nabla \phi(x), \\
 \nabla\Phi(u^{-1}(x)) &= \nabla \phi(x).
\end{align*} 
With this equality and the definition for $u^{-1}(x)$ in Eq. \ref{eq:g}, one 
can now verify that the following equality holds
\begin{align} 
    x &= u^{-1}(x) - \frac{\lambda z}{2\pi} \nabla \Phi (u^{-1}(x)).
    \label{eq:blahblah}
\end{align} 
But since $u(u^{-1}(x)) = x$ we can identify $u(x)$ with
\begin{align} 
    u(x) &= x - \frac{\lambda z}{2\pi}\nabla \Phi(x).
    \label{eq:u}
\end{align} 
%
Furthermore, we can evaluate $z(u(x))$ in terms of $\Phi(x)$ by making use of 
the following relation
\begin{align*} 
    \frac{\partial}{\partial x} u^{-1}(x) &= \left[1 + \frac{\lambda z}{2\pi} 
    \nabla^2 \phi(x)\right] = \frac{z}{z(x)},
\end{align*}
where we have used the expression for $z(x)$ in Eq. \ref{eq:zx}, then taking 
the derivative with respect to $x$ of both sides of Eq. \ref{eq:blahblah} 
\begin{align*} 
 \frac{\partial}{\partial x}x = 1 &= \frac{\partial}{\partial x} 
 \left[u^{-1}(x) - \frac{\lambda z}{2\pi} \nabla \Phi (u^{-1}(x))\right], \\
 &= \frac{z}{z(x)} - \frac{z}{z(x)} \frac{\lambda z}{2\pi} 
    \nabla^2 \Phi (u^{-1}(x)) 
\end{align*}
and rearranging
\begin{align} 
    z_\Phi(x) &\equiv z(u(x)) = z \left[ 1 - \frac{\lambda z}{2\pi} 
    \nabla^2 \Phi(x) \right] 
    \label{eq:zPhi}
\end{align} 
Inserting Eqs \ref{eq:pyx}, \ref{eq:u} and \ref{eq:zPhi} into Eq. \ref{eq:Fr2coh} 
we can then write
\begin{align} \label{eq:Fr5}
    \psi(x, z) &\approx e^{-2\pi i z_\Phi(x)/\lambda} \times \nonumber \\
    &\qquad\qquad p(x, z) \psi_\text{ref}(x - \frac{\lambda z}{2\pi} \nabla \Phi(x), z_\Phi(x)).
\end{align}
%
%
%
or, the inverse relationship:
\begin{align} \label{eq:Fr6}
    p(x, 0) \psi_\text{ref}(x, z(x)) &\approx e^{-2\pi i (z-z(x))/\lambda} \times \nonumber \\
    &\qquad\qquad \sqrt{\frac{z}{z(x)}} \psi(x + \frac{\lambda z}{2\pi} \nabla \phi(x), z).
\end{align}
%
    %
%
This formulation of the projected image separates the effects of the geometric 
and propagation based distortions induced in the detected image by phase 
variations in the incident illumination. The geometric distortions are captured 
by the term $\lambda z/ (2 \pi) \nabla \Phi(x)$ and the change in the fringe 
structure of a feature by the term $z_\Phi(x)$.

For the purposes of the current work, we will ignore variations in the fringing 
terms $z(x)$ and $z_\Phi(x)$ across the wavefield and use instead the constants
\begin{align}
  z_\Phi(x) &\approx \bar{z}_\Phi \equiv z\left[1 - 
  \frac{\lambda z}{2\pi} \langle\nabla^2 \Phi(x)\rangle_{x}\right], \\
  z(x) &\approx \bar{z} \equiv \left[\frac{1}{z} - 
  \frac{\lambda}{2\pi} \langle\nabla^2 \phi(x)\rangle_{x}\right]^{-1}. 
  \label{eq:zbar}
\end{align}
$\langle\nabla^2 \phi(x)\rangle_{x}$ and $\langle\nabla^2 \Phi(x)\rangle_{x}$ 
are the mean phase curvatures of the illumination in the plane of the sample 
and the detector respectively. Under the Fresnel approximation, they can be 
defined in terms of the effective source distance from the sample or detector
planes. If $z_1$ is the distance between the entrance surface of the sample and
the effective source point, then
\begin{align*}
 \langle\nabla^2 \phi(x)\rangle_{x} &= \frac{2\pi}{\lambda z_1} & 
 \langle\nabla^2 \Phi(x)\rangle_{x} &= \frac{2\pi}{\lambda (z+z_1)}
\end{align*}
Using the above expressions, one can now verify that
\begin{align}
  \bar{z} = \bar{z}_\Phi &= \frac{z z_1}{(z+z_1)}.
\end{align}
With the above approximations for $z(x)$ and $z_\Phi(x)$ and taking the mod square of Eqs. \ref{eq:Fr5} and \ref{eq:Fr6}, we arrive at the 1D speckle tracking approximation
\begin{align} 
    I(x, z) &\approx W(x) I_\text{ref}(x - \frac{\lambda z}{2\pi} \nabla \Phi(x), 
                     \bar{z}),  \\
    w(x) I_\text{ref}(x, \bar{z}) &\approx \frac{z}{\bar{z}} I(x + \frac{\lambda z}{2\pi} 
                                \nabla \phi(x), z).
\end{align}

In 2D, the constant prefactor to the Fresnel integral in Eq. \ref{eq:Fcoh} is $1/(-i\lambda z)$ (rather than $1/\sqrt{-i\lambda z}$ in 1D). This leads to an altered expression for $W$
\begin{align}
 W(\textbf{x}) &\equiv \left(\frac{\bar{z}}{z}\right)^2 w(\textbf{x}).
\end{align}
Additionally, $\bar{z}$ is now defined by the average wavefront curvature over the 2D transverse plane:
\begin{align*}
 \bar{z} \equiv z\left[1 - 
  \frac{\lambda z}{2\pi} \frac{1}{2} \langle \nabla^2 \Phi(\textbf{x}) \rangle_{\textbf{x}}\right]
\end{align*}
Where the Laplacian operator is now also over the 2D plane $\nabla^2 \equiv \frac{\partial^2 }{\partial x^2} + \frac{\partial^2 }{\partial y^2}$. If one accepts the approximation $z(\textbf{x}) \approx \bar{z}$ from the beginning, then the analysis presented here in 1D can be repeated in 2D by following the above steps, first along the x-axis and then along the y-axis. The result of this procedure is
\begin{align} 
    I(\textbf{x}, z) &\approx W(\textbf{x}) I_\text{ref}(\textbf{x} - \frac{\lambda z}{2\pi} \nabla \Phi(\textbf{x}), 
                     \bar{z}), \\
    w(\textbf{x}) I_\text{ref}(\textbf{x}, \bar{z}) &\approx \left(\frac{z}{\bar{z}}\right)^2 I(\textbf{x} + \frac{\lambda z}{2\pi} 
                                \nabla \phi(\textbf{x}), z).
\end{align}

\section{Derivation of the Limits to the Approximation}\label{sec:limits_der}

In section \ref{sec:approx} of the appendix, it was shown that the Fresnel integral of $T(x)$, modulated by $p(x, 0)$\footnote{For simplicity, the following analysis will be carried out in 1D. Generalisation to 2D is not required to support the results of this section.} :
\begin{align}\label{eq:aF}
 \psi(x, z) &= \frac{e^{2\pi i z/\lambda}}{\sqrt{-i\lambda z}} \int p(x', 0) T(x') e^{\frac{i\pi}{\lambda z}(x-x')^2} dx',
\end{align}
can be approximated by
\begin{align}\label{eq:aF2}
 \psi(x, z) &\approx \frac{p(x, z)}{\sqrt{-i\lambda \bar{z}}} \int T(x') e^{\frac{i\pi}{\lambda \bar{z}}(x'-u(x))^2} dx',
\end{align}
subject to the approximations
\begin{align*}
 &\text{1:} &g(x, x') &\approx g(x, u(x)) + 
             \frac{1}{2} \bigg[ \nabla^2\phi(u(x)) + 
             \frac{2\pi}{\lambda z}\bigg] (x'-u(x))^2, \\
 &\text{2:} &\sqrt{w}(x') &\approx \sqrt{w}(u(x)), \\
 &\text{3:} &z(x) &\approx \bar{z} \equiv \frac{z z_1}{z+z_1},  
\end{align*}
where these are additional to the approximations necessary for the paraxial approximation to hold, $p(x, 0) = \sqrt{w}(x) e^{i\phi(x)}$, $p(x, z) = e^{2\pi i z/\lambda} \sqrt{W}(x) e^{i\Phi(x)}$, $u(x) = x - \frac{\lambda z}{2\pi} \nabla \Phi(x)$, $z_1$ is the effective distance between the light source and the sample plane and $g(x, x') \equiv \pi\frac{(x-x')^2}{\lambda z} + \phi(x')$.

In this section we shall determine the requirements for each of these three approximations to hold.  

\subsubsection{The First Approximation}

Let us assume for the moment that approximations 2 and 3 are valid. In this case, it is sufficient to demand that the Taylor series expansion of $g(x, x')$ is valid within the interval of convergence of the Fresnel integral in Eq. \ref{eq:aF2}. 

In appendix \ref{sec:approx}, we have intuited that the expansion of $g(x, x')$ need only be valid for values of $x'$ satisfying $q_g(x, x') < q_T$, where $q_g(x, x')$ are the spatial frequencies of $g$ and $q_T$ is the maximum spatial frequency of $T$. For values of $x'$ outside of this region, successive oscillations of the integrand will occur at roughly the same magnitude ($\pm T(x')$) with a net zero contribution to the integral.

So, let us first examine the domain of $x'$ over which the expansion is valid. The Lagrange error bound for the Taylor series expansion of $g(x, x')$, about $x' = u(x)$, sets a limit on the magnitude of the residual:
\begin{align*}
 |R_N(x'; g)| &= |g(x, x') - g_N(x, x')| \\
            &\le \bigg| \frac{M}{(N+1)!} (x'-u(x))^{N+1}\bigg|\\
           \text{where } |\frac{\partial^{N+1} g}{\partial x'^{N+1}}(x, x'')| &\le M \text{ for } |x''-u(x)| < |x'-u(x)|
\end{align*}
With the above expression, we can identify the interval $2d_\phi$ of $x'$ about $u(x)$, such that $x': u(x) - d_\phi \rightarrow u(x)+d_\phi$, for which the magnitude of the residual $|R_N|$ is below some threshold level of tolerance. So, for $N=2$, 
\begin{align}
 \frac{\partial^3 g}{\partial x'^3}(x, x') &= \phi^{(3)}(x') \text{, and } \\
 \phi^{(3)}_\text{max} &\equiv M = \text{max}\left( |\phi^{(3)}(x')| : |x'-u(x)| \le d_\phi \right),
\end{align}
we have:
\begin{align}\label{eq:dd}
 E_2 &= \frac{1}{6} \phi^{(3)}_\text{max} d_\phi^3 < \frac{1}{6} E_\text{tol},  \nonumber \\
 \text{ or } d_\phi &< \left(\frac{E_\text{tol}}{\phi^{(3)}_\text{max}}\right)^\frac{1}{3},
\end{align}
where $E_2$ is the Lagrange error bound for the second-order expansion of $g$ and $E_\text{tol}$ is maximum tolerable error (which has been implicitly defined so as to absorb the factor of 6). 

Now we determine the minimum interval $2d_{q_T}$ required for the Fresnel integral in Eq. \ref{eq:aF2} to converge near to its true value. 

Let us expand $T(x')$ as Fourier series, such that $T(x') = \int \hat{T}(q) e^{2\pi i x' \cdot q}dq$, where $\hat{T}(q)$ is the complex amplitude of the for the $q$\textsuperscript{th} full-period spatial frequency of $T(x')$ corresponding to features of extent $X=1/q$. 

By the superposition principle, we can select the highest spatial frequency of $T$ such that $\hat{T}(q) = 0$ for $|q|>q_T$ and examine the radius of convergence of the integral in Eq. \ref{eq:aF2} for $T(x') \rightarrow e^{2\pi i x' \cdot q_T}$:
\begin{align}
 \psi(x, z; q_T) &\equiv p(x, z) \int_{u(x)-d(q_T)}^{u(x)+d(q_T)} e^{2\pi i x' \cdot q_T} e^{i\pi\frac{(x'-u(x))^2}{\lambda \bar{z}}} dx'.
\end{align}
This integral can reduced to the following form:
\begin{align}
 \psi(x, z; q_T) &= \cdots  \int_{\sqrt{\frac{2}{\lambda \bar{z}}}(\lambda \bar{z} q_T-d)}^{\sqrt{\frac{2}{\lambda \bar{z}}}(\lambda \bar{z} q_T+d)} e^{i\frac{\pi}{2} v^2} dv \nonumber \\
 &= \cdots \bigg[ E\left( \sqrt{\frac{2}{\lambda \bar{z}}}(\lambda \bar{z} q_T+d) \right) \nonumber \\
 &\qquad - E\left( \sqrt{\frac{2}{\lambda \bar{z}}}(\lambda \bar{z} q_T-d) \right)\bigg] 
 \label{eq:E}
\end{align}
where the function $E$ is known as the Euler or Cornu spiral and the ``$\cdots$'' represent terms that are constant with respect to the integration variable $v$. The Euler spiral can be constructed in the complex plane in terms of the Fresnel integrals $C$ and $S$:
\begin{align}
 E\left( x \right) &= C(x) + iS(x)\\
 &= \int_0^x \cos(\frac{\pi}{2}x^2) dx + i\int_0^x \sin(\frac{\pi}{2}x^2) dx.
\end{align}
As both $C$ and $S$ approach their limit of $\frac{1}{2}$ as $x\rightarrow \infty$ with equal rapidity, we shall examine only the imaginary part of $E$ for convenience. 
\begin{figure}
\includegraphics[width=8.88cm]{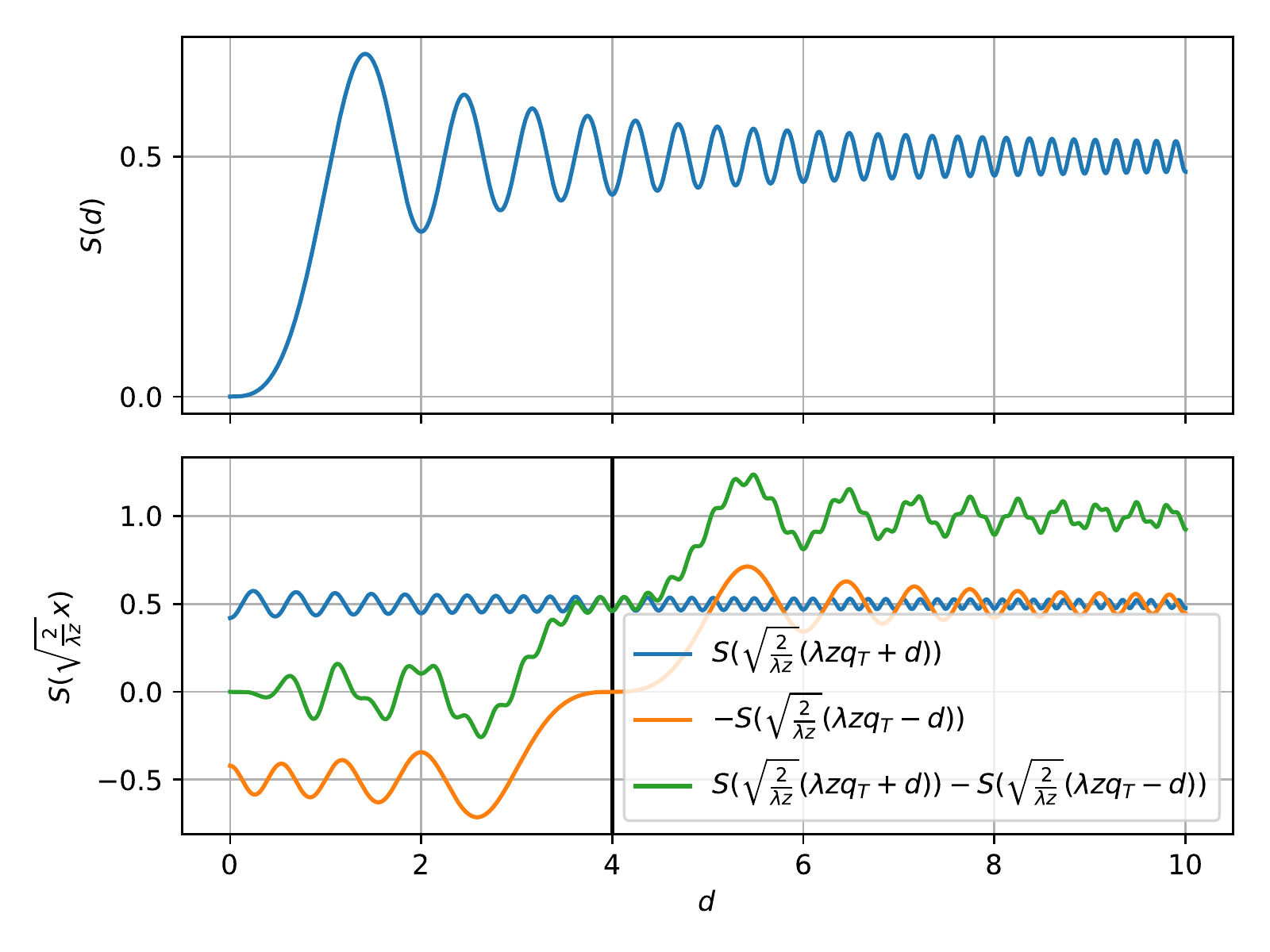}
\caption{\textbf{First row}: The Fresnel integral $S$. \textbf{Second row}: The terms $\propto \psi(x,z; q_T)$ in Eq \ref{eq:E}. For these plots, $\lambda \bar{z}= 2$ and the vertical black line is at $d=\lambda \bar{z} q_T$. } 
\label{fig:Euler}
\end{figure}
In Fig. \ref{fig:Euler} we plot the imaginary part of the indefinite integral of Eq. \ref{eq:E} (shown in green in the second row) as a function of $d$. We can see here that for $d\gtrapprox\lambda \bar{z} q_T$ the integral oscillates about $1$ and that the amplitude of these oscillations are dominated by the $S(\sqrt{2/\lambda \bar{z}}(\lambda \bar{z} q_T - d))$ term (shown in orange). 

The extrema of $S(x)$ are given by:
\begin{align}
   \frac{\partial S(x)}{\partial x} &= \sin(\frac{\pi}{2}x^2) = 0 \\
   \therefore x_m &= \pm \sqrt{2m}.
\end{align}
The extrema of $S(\sqrt{2/\lambda \bar{z}}(\lambda \bar{z} q_T - d))$ are therefore located at:
\begin{align}\label{eq:d}
 d_m &= \sqrt{m\lambda \bar{z}} + \lambda \bar{z} q_T \text{ for } m=0,1,2, \cdots 
\end{align}
So, the residual for the integral in Eq \ref{eq:E} is proportional to $1/2 - S(\sqrt{2m})$. Therefore the integer $m$ will serve as a measure of convergence for the integral and we can demand that $d_{q_T} > d_m$. Finally, combining Eqs \ref{eq:dd} and \ref{eq:d} we have, for $d_{q_T} < d_\phi$:
\begin{align}\label{eq:blah}
 \sqrt{m\lambda \bar{z}} + \lambda \bar{z} q_T &< \left(\frac{E_\text{tol}}{\phi^{(3)}_\text{max}}\right)^\frac{1}{3},
\end{align}
The above inequality can be seen as a limit for the phase variation $\phi^{(3)}_\text{max}$ or, for a given $\phi^{(3)}_\text{max}$, as a limit on the smallest features that will be resolved ($q_T = 1/X$) according to the approximation in Eq. \ref{eq:aF2}. Setting, $\sqrt{m}=1$, multiplying both sides of Eq. \ref{eq:blah} by $\phi^{(3)}_\text{max}$ and raising both sides by the third power, we have the condition:
\begin{align}\label{eq:blah2}
 \frac{(\sqrt{\lambda \bar{z}} + \lambda \bar{z} q_T)^3}{\phi^{(3)}_\text{max}} &\ll 1,
\end{align}
%

%
%
%
%

\subsubsection{Approximation 2}

Assuming for now that approximations 1 and 3 are valid, we require that $\sqrt{w}(x') \approx \sqrt{w}(u(x))$ is valid within the interval $|x' - u(x)| < d_w$, and that this interval is larger that $d_{q_T}$, which is the interval about $x'=u(x)$ over which the integral in Eq \ref{eq:aF2} will converge. Making use, once again, of the Lagrange error bound we have:
\begin{align}
 d_w &< \frac{E_\text{tol}}{\sqrt{w}^{(1)}_\text{max}}.
\end{align}
The requirement $d_w > d_{q_T}$ leads to:
\begin{align} \label{eq:dwdq}
 \frac{\sqrt{\lambda \bar{z}} + \lambda \bar{z} q_T}{\sqrt{w}^{(1)}_\text{max}} &\ll 1.  
\end{align}
%
%
%


\subsubsection{Approximation 3}
Assuming once again that approximations 1 and 2 are valid, the governing equation becomes:
\begin{align}
 \psi(u^{-1}(x), z) &\approx \frac{p(u^{-1}(x), z)}{\sqrt{-i\lambda \bar{z}}} \int T(x') e^{\frac{i\pi}{\lambda z_\Phi(x)}(x'-x)^2} dx',
\end{align}
where we have used $u^{-1}(x) = x + \frac{\lambda z}{2\pi}\nabla \phi(x)$ and restored $z_\Phi(x)$ in place of $\bar{z}$ in Eq. \ref{eq:aF2}. This approximation, that $z(x) \approx z_\Phi(x) \approx \bar{z}$, is thus valid in the limit where the residual term:
\begin{align}
R &= \frac{p(u^{-1}(x), z)}{\sqrt{-i\lambda \bar{z}}} \int T(x') \left[ e^{\frac{i\pi}{\lambda \bar{z}} (x'-x)^2} - e^{\frac{i\pi}{\lambda z_\Phi(x)} (x'-x)^2} \right] dx', \\
&= \psi(u^{-1}(x), \bar{z}) - \psi(u^{-1}(x), z_\Phi(x)),
\end{align}
approaches zero. The conditions under which $\psi(x, z_1) \approx \psi(x, z_2)$, in the Fresnel diffraction regime, are well known: the requirement is that the Fresnel number $F = X^2 / \lambda |z_2-z_1| \gg 1$. In this case, $|z_1-z_2| = |z_\Phi(x) - \bar{z}|$ which depends on $x$. To generalise this requirement across the entire wavefront, we thus demand that:
\begin{align}
 \frac{X^2}{\lambda \sigma(z)} &\gg 1, \\
 \text{ or } F &\gg \frac{\sigma(z)}{\bar{z}},
\end{align}
where $\sigma(z)$ is the standard deviation of the effective propagation distance given by $\sigma(z) \equiv \sqrt{ \langle (z_\Phi(x) - \bar{z})^2 \rangle_x }$ and $F = X^2/\lambda \bar{z}$ is the Fresnel number for features of size $X$ propagating a distance $\bar{z}$. Under this condition, features of size $X$ (after correcting for the geometric distortions) will produce the same image on the detector regardless of their transverse position along the wavefront. 

This condition can be expressed as a constraint on the phase profile of the beam. Using the definitions for $z_\Phi(x)$ (Eq. \ref{eq:zPhi}) and $\bar{z}$ (Eq. \ref{eq:zbar}) we have:
\begin{align}
 z_\Phi(x) - \bar{z} &= \frac{\lambda z^2}{2\pi}\left( \nabla^2 \Phi(x) - \langle \nabla^2 \Phi(x) \rangle_x \right),  \\
  \sigma(z) &= \frac{\lambda z^2}{2\pi} \sqrt{ \left( \nabla^2 \Phi(x) - \langle \nabla^2 \Phi(x) \rangle_x \right)^2 } \\
  &= \frac{\lambda z^2}{2\pi} \sigma(\Phi^{(2)}). 
\end{align}
Using the above equation we have that approximation 3 is valid in the limit:
\begin{align}
  F \gg \frac{\lambda z^2}{2\pi \bar{z}} \sigma(\Phi^{(2)}).
\end{align}

\subsection{Limit on the defocus for an ideal lens}
The illumination formed by an ideal lens, with a hard edged aperture, has a distinct form, see for example Fig. \ref{fig:stem}. Within the plateau of the wavefront the intensity oscillates about a mean value, while the phase profile is approximately quadratic $\phi \approx \pi x^2/\lambda z_1$. In this case, approximation 2, that $\sqrt{w}(x') \approx \sqrt{w}(u(x))$, is the most onerous of the three. In the above analysis, we had used the Lagrange error bound to estimate the maximum distance along the wavefront ($|x'-x|$) for which this approximation will hold. But for the present case, this estimate is over-bounded given its general nature. Instead, we propose that an acceptable condition for this approximation is that $\sqrt{w}(x') \approx \sqrt{w}(u(x))$ will remain valid for $|x'-x| < z_1 \text{NA}$, where $\text{NA}$ is the numerical aperture of the lens and $z_1 \text{NA}$ is approximately equal to the half width of the plateau. Thus we replace $d_w = E_\text{tol}/\sqrt{w}^{(1)}_\text{max}$ with $d_w = z_1 \text{NA}$ and Eq. \ref{eq:dwdq} becomes:
\begin{align} \label{eq:dwdq_stem}
 \frac{\sqrt{m\lambda \bar{z}} + \lambda \bar{z} q_T}{z_1 \text{NA}} &\ll 1 ,  
\end{align}
where this condition applies for features within the plateau of the illumination. 
%
%
\end{document}